\documentclass[prd,twocolumn,showpacs,reprint,preprintnumbers,nofootinbib,amsmath,amssymb]{revtex4-2}

\RequirePackage[colorlinks=true
,urlcolor=blue
,anchorcolor=blue
,citecolor=blue
,filecolor=blue
,linkcolor=blue
,menucolor=blue
,linktocpage=true
,pdfproducer=medialab
,pdfa=true
]{hyperref}

\usepackage{amsmath,amssymb,amsthm,amsfonts}
\usepackage{graphicx,tabularx}
\usepackage{color}
\usepackage{multirow}  
\usepackage{comment}
\usepackage{slashed}
\usepackage{enumitem}
\usepackage{cleveref}
\usepackage{rotating}
\usepackage[english]{babel}
\usepackage[justification=centerlast]{caption}
\usepackage{subcaption}
\usepackage{float}

\usepackage{breqn} 
\makeatletter 
\let\cref@old@eq@setnumber\eq@setnumber 
\def\eq@setnumber{%
\cref@old@eq@setnumber%
\cref@constructprefix{equation}{\cref@result}%
\protected@xdef\cref@currentlabel{%
[equation][\arabic{equation}][\cref@result]\p@equation\theequation}} 
\makeatother

\usepackage{cases}

\crefname{section}{Sec.}{Secs.}
\crefname{figure}{Fig.}{Figs.}
\crefname{equation}{Eq.}{Eqs.}
\crefname{appendix}{Appendix}{Appendices}
\setlist[description]{leftmargin=0.4cm}
\setlist[itemize]{leftmargin=0.4cm}

\newcommand{\be}{\begin{equation}\begin{aligned}}
\newcommand{\ee}{\end{aligned}\end{equation}}

\newcommand{\beq}{\begin{equation}}
\newcommand{\eeq}{\end{equation}}
\newcommand{\beqa}{\begin{eqnarray}}
\newcommand{\eeqa}{\end{eqnarray}}

\newcommand{\mev}{\text{MeV}}
\newcommand{\gev}{\text{GeV}}
\newcommand{\tev}{\text{TeV}}

\newcommand{\cm}{\text{cm}}
\newcommand{\m}{\text{m}}

\newcommand{\km}{\text{km}}

\renewcommand{\eqref}[1]{Eq.~(\ref{#1})}

\newcommand{\eg}{{\em e.g.}}

\newcommand{\TODO}[1]{\textcolor{green}{TODO}}
\RequirePackage[normalem]{ulem}

\DeclareUnicodeCharacter{2212}{\textendash}

\makeatletter
\def\l@subsubsection#1#2{}
\makeatother

\usepackage{multirow}
\usepackage{makecell}

\usepackage{fontawesome}

\begin{document}

\title{Probing some photon portals to new physics at intensity frontier experiments}

\author{Krzysztof Jod\l{}owski}
\email{k.jodlowski@ibs.re.kr}
\affiliation{Particle Theory and Cosmology Group\char`,{} Center for Theoretical Physics of the Universe\char`,{} Institute for Basic Science (IBS)\char`,{} Daejeon\char`,{} 34126\char`,{} Korea}

\begin{abstract}
A number of extensions of the Standard Model predict the existence of new light, weakly-coupled particles that couple to the visible sector through higher-dimensional operators containing one or two photons, suppressed by a high new physics scale, and thus have long lifetimes.
In this work, we study the prospects for detecting three $\sim\,$sub-GeV such long-lived particles (LLP) at intensity frontier experiments: a massive spin-2 mediator ($G$), a dark axion portal, and a light neutralino coupled to ALPino or gravitino. 
We consider the production and visible decays of these particles in several current and proposed beam dump experiments (CHARM, NuCal, SeaQuest, NA62, SHiP) as well as in the LHC detectors (FASER, FASER$\nu$, FLArE, MATHUSLA).
In addition to the usual displaced vertex signature, we also examine the impact of electron scattering signature and the Primakoff-like process which leads to conversion of $G$ into a photon or to the secondary LLP production via upscattering of the lighter dark sector state on dense material put in front of the detector.
In all cases, we find the SHiP experiment could provide the strongest constraints for the displaced vertex search, while FASER2/FPF could provide complementary coverage of the $\gamma c\tau \sim 1\,\m$ decay length region of the parameter space.
\href{https://github.com/krzysztofjjodlowski/Looking_forward_to_photon_coupled_LLPs}{\faGithub} 
\end{abstract}

\renewcommand{\baselinestretch}{0.85}\normalsize
\maketitle
\renewcommand{\baselinestretch}{1.0}\normalsize

\section{\label{sec:intro}Introduction}
Thanks to various observations of the Universe, it has been established that non-baryonic dark matter (DM) outweighs ordinary matter described by the Standard Model (SM) of fundamental interactions by about five to one \cite{Einasto:2009zd,Bertone:2016nfn}.
Despite intensive searches for leading DM candidates such as axions \cite{Marsh:2015xka,Choi:2020rgn} and Weakly Interacting Massive Particles (WIMPs) \cite{Roszkowski:2017nbc,Arcadi:2017kky}, no clear signal has been found so far.
This motivates exploring alternative scenarios \cite{Bertone:2018krk}, opening up a much broader range of DM candidates leading to unique observational signatures.

One promising possibility is that DM is part of a larger landscape of many new particles and interactions, the so-called Dark Sector (DS) \cite{Essig:2013lka,Alexander:2016aln}.
The weak connection between the visible and invisible sectors is described by an effective Lagrangian containing operators ordered by their mass dimension often called portals or simplified models.
The renormalizable ones are: scalar (dark Higgs) portal \cite{Patt:2006fw,Arcadi:2019lka}, vector (dark photon) portal \cite{Okun:1982xi,Holdom:1985ag}, and sterile neutrino portal \cite{Weinberg:1979sa,Bondarenko:2018ptm}, where the portal name indicates the spin of the mediator.

To satisfy experimental constraints, the dimensionless couplings of these operators must be very small, while higher dimensional operators need to be suppressed by an appropriate power of sufficiently large $\Lambda$, the scale of new physics.
Such very weak couplings typically result in making the mediator a long-lived particle (LLP) - $c\tau \gtrsim 1\m$ - which makes them the prime target for intensity frontier experiments searching for $\sim\,$sub-GeV, highly boosted, and very weakly coupled Beyond the Standard Model (BSM) species \cite{Essig:2013lka,Battaglieri:2017aum}.

At mass-dimension 5, the most studied operator is arguably the axion portal \cite{Peccei:1977hh} - which in its minimal form couples an axion-like particle (ALP) to two photons. Such a coupling leads to a diverse and interesting phenomenology, and due to its suppressed decay width for $\sim\,$sub-GeV ALP, is one of the main benchmarks of current and future intensity frontier searches \cite{Battaglieri:2017aum,Beacham:2019nyx}.

In this work, we explore the prospects of intensity frontier searches for other well-motivated LLPs coupled to photons by higher dimensional operators: a massive spin-2 mediator ($G$), which is characterized by a two-photon coupling, and several scenarios in which the dark sector states interact only with a single photon. The latter models include the dark axion portal, where either a scalar ALP or a vector dark photon (DP) can act as a LLP, or a light neutralino coupled to ALPino or gravitino.

The massive graviton-like mediator has recently emerged as an interesting portal between the SM and the DS, \eg, in the context of ``Gravity-mediated dark matter" \cite{Lee:2013bua,Lee:2014caa,Bernal:2018qlk,Kang:2020huh}, theories with extra spatial dimensions \cite{Rueter:2017nbk,Folgado:2019sgz,Folgado:2019gie}, and within other similar frameworks \cite{Bernal:2018qlk,Kraml:2017atm,Rueter:2017nbk,Cai:2021nmk}.
Searches for $\sim\,$sub-GeV massive spin-2 mediator has been considered previously in \cite{Kang:2020huh,Voronchikhin:2022rwc,Voronchikhin:2023znz}, where significant constraints on the mass and the coupling strength of $G$ were obtained by considering the missing energy signature in electron fixed target experiments. 
However, visible decays of $G$ have not been considered before. They lead to a displaced vertex signature which can be searched at the intensity frontier experiments \cite{Beacham:2019nyx} with almost zero background. 
We fill this gap by simulating $G$ production and decay in a number of past and upcoming beam dump and LHC experiments. 
We also investigate the signature of a single high-energy photon produced by the conversion process $G \to \gamma$ taking place by a Primakoff-like scattering at the future LHC experiments such as FASER$\nu$2 \cite{Batell:2021blf,Anchordoqui:2021ghd} and FLArE \cite{Batell:2021blf,Kling:2022ehv}.

The dark axion portal (DAP) is characterized as a dim-5 interaction between an ALP, a DP, and photon induced by interactions in the DS \cite{Kaneta:2016wvf,Ejlli:2016asd}.  
Such a mechanism can take place, \eg, due to 1-loop processes involving massive dark fermions charged under global Peccei-Quinn symmetry $U(1)_{\text{PQ}}$ and gauge groups - $U(1)_{\text{Y}}$ (hypercharge), and $U(1)_{\text{Dark}}$ - which can be viewed as a generalization of the KSVZ \cite{Kim:1979if,Shifman:1979if} axion to DS containing new $U(1)_{\text{Dark}}$ gauge group.
For this model, we extend the previous works \cite{deNiverville:2018hrc,deNiverville:2019xsx} in several directions. 
We re-examine the LLPs production mechanisms and find that the previously neglected vector meson decays actually provide the leading contributions to the LLP yields, allowing much larger coverage of the parameter space than previously found for past and future beam dump and LHC experiments.
We also take advantage of the recent developments concerning the FASER experiment, which began collecting data at the start of Run 3 of the LHC. In particular, a dedicated neutrino emulsion detector FASER$\nu$ \cite{FASER:2019dxq,FASER:2020gpr} has been installed in front of the main detector. It is made of tungsten layers, therefore, it can act as a target for the secondary production of LLPs by Primakoff-like upscattering; see \cite{Jodlowski:2019ycu,Jodlowski:2020vhr} for studies dedicated to the non-minimal scalar, vector, and sterile neutrino portals.
Moreover, FASER2 will be sensitive to semi-visible LLP decays depositing energy in the calorimeter only via a single high-energy photon \cite{Boyd:2803084,Jodlowski:2020vhr}.

Such a signature can also help constrain scenarios of low-energy supersymmetry (SUSY) \cite{Golfand:1971iw} breaking, in particular the displaced semi-visible decays of bino.
It is well known that unstable neutralino could be very light, possibly with masses in the sub-GeV range \cite{Gogoladze:2002xp,Dreiner:2009ic}, provided it is predominantly composed of bino, since then its couplings to gauge bosons vanish and stringent collider constraints are relaxed.
While recent studies \cite{Gorbunov:2015mba,Dercks:2018eua,Choi:2019pos} have investigated such light binos in R-parity violating SUSY scenarios, we consider two alternative scenarios that preserve R-parity: bino coupled to ALPino or gravitino and a photon.
The first model contains a SUSY partner of an ALP called ALPino, while the second one is based on local SUSY, which predicts a spin-$3/2$ SUSY partner of a graviton called gravitino.
Then, the relevant coupling to a photon is proportional to the Peccei-Quinn (PQ) or SUSY breaking scale, respectively. As we shall show, the displaced vertex search at beam dumps will allow one to provide complementary coverage of the parameter space or to even improve the existing bounds for such sub-GeV bino.

The paper is organized as follows. 
In \cref{sec:models} we introduce the BSM scenarios we investigate further.
We discuss their main physical properties in the context of long-lived regime relevant for the displaced vertex searches.
In \cref{sec:LLP_prod_and_sign} we describe the main experimental signatures of LLPs. We also discuss the framework of our simulations.
In \cref{sec:results} we present and discuss our main results - projections of the sensitivities of future experiments looking for LLPs.
In \cref{sec:conclusions} we summarize our findings, while technical details of the analysis are presented in \cref{app:decays,app:prod_vec,app:d2Br,app:sigma_Prim}.
Finally, our simulation is implemented within an extended version of the \texttt{\href{https://github.com/KlingFelix/FORESEE}{FORESEE}} \cite{Kling:2021fwx} package, which can be found in \href{https://github.com/krzysztofjjodlowski/Looking_forward_to_photon_coupled_LLPs}{\faGithub}.

\section{Models\label{sec:models}}

\subsection{Massive spin-2 mediator\label{sec:massive_spin_2}}
The massive spin-2 mediator couples to the energy-momentum tensor of gauge and/or matter fields, and hence is described by the following effective Lagrangian in the electromagnetic (EM) sector \cite{Han:1998sg,Giudice:1998ck,Lee:2013bua}:\footnote{We leave the general case for further study.}
\begin{dmath}[labelprefix={eq:}]
  \mathcal{L} \supset g_{\gamma\gamma}\, G^{\mu \nu}\left(\frac{1}{4} \eta_{\mu \nu} F_{\lambda \rho} F^{\lambda \rho}+F_{\mu \lambda} F_\nu^{\ \lambda}\right) -i \sum_l \frac{g_l}{2} G^{\mu \nu}\left(\bar{l} \gamma_\mu D_\nu l-\eta_{\mu \nu} \bar{l} \gamma_\rho D^\rho l\right),
  \label{eq:L_G2}
\end{dmath}
where $g_{\gamma\gamma}$ and $g_l$ are mass-dimension -1 couplings, $F_{\mu\nu}$ is the EM field strength tensor, and $G_{\mu \nu}$ describes the massive graviton field corresponding to the perturbation of a given metric tensor $g_{\mu\nu}$ around the Minkowski metric $\eta_{\mu\nu}$: $g_{\mu\nu} \approx \eta_{\mu\nu} + 2/m_{\mathrm{Pl. red.}} G_{\mu \nu}$ \cite{Veltman:1975vx}, where $m_{\mathrm{Pl. red.}} = 2.4 \times 10^{18}\,\gev$ is the reduced Planck mass.

Since an axion-like particle (ALP) coupled only to two photons or only to leptons is one of the main benchmarks for intensity frontier searches, by analogy, one might want to restrict the general case of massive spin-2 portal described by \cref{eq:L_G2} to the cases: $g_{\gamma\gamma} \neq 0$ or $g_l \neq 0$, while neglecting the other coupling.
However, as discussed in \cite{Artoisenet:2013puc}, such interactions lead to perturbative unitarity violation in the $m_G \to 0$ regime for the $q \bar{q} \to G g$ process (quark-antiquark annihilation to $G$ and a gluon), unless the couplings have universal form, $g_l=g_{\gamma\gamma}$.
In fact, for non-universal couplings, the cross-section for this process scales like $\propto 1/m_G^4$, due to lack of decoupling of the helicity-0 modes of $G$. What is more, the helicity-1 modes lead to $\propto 1/m_G^2$ dependence, while only the helicity-2 modes are free from such an enhancement.
Recent work \cite{Cai:2021nmk} has considered the freeze-in of sub-MeV massive spin-2 graviton produced by the process considered in \cite{Artoisenet:2013puc}, claiming, they found that in the small mass regime it behaves as $\propto s/m_G^4$, where $s$ is the square of the center-of-mass energy. 
Although follow-up work \cite{Gill:2023kyz} devoted to an explicit calculation of such a process again found that the $m_G \to 0$ limit for universal coupling is finite, both the universal coupling interactions and the coupling to a pair of photons only are interesting, and can be easily constructed from extra-dimensional theory, therefore we consider both of them.

Previous works \cite{Kang:2020huh,Voronchikhin:2022rwc,Voronchikhin:2023znz} have focused on the invisibly decaying massive spin-2 mediator, which was constrained by the missing energy searches at BaBar \cite{BaBar:2001yhh} and NA64e \cite{Banerjee:2019pds}, and will be used by the next-generation experiments such as NA64$\mu$ \cite{Sieber:2021fue}, LDMX \cite{Mans:2017vej}, and M$^3$ \cite{Kahn:2018cqs}.
It is well-known that the visible decays of LLPs with middle range decay lengths, $10\,\m \lesssim d \lesssim 10\,\km$, provide one of the strongest constraints on such particles \cite{Battaglieri:2017aum,Beacham:2019nyx}. 
Therefore, we investigate both scenarios of $G$ coupling by its decays into a pair of photons or charged leptons, and by the $G \to \gamma$ conversion \cite{Kling:2022ehv}.

The lifetime of $G$ depends on the widths of the two-body decays, and the typical decay length of $G$ that can be probed at beam dumps is
\be
  d_{G} \simeq &\, 100 \,\m \times \left(\frac{E}{1000\,\gev}\right) \left(\frac{0.1\,\gev}{m_G}\right)^4 \left(\frac{5.75 \times 10^{-5}}{g_{\gamma\gamma}}\right)^2,
  \label{eq:ctau_G2_universal}
\ee
where $d_{G} = c \tau_G \beta \gamma$, $\gamma = E/m_G$ is the boost factor of $G$ in the LAB frame, $\beta = \sqrt{1-1/\gamma^2}$, and $\tau_G = 1 / \Gamma_{G}$, given by \cref{eq:Gamma_G2}.
In the case of non-universal coupling the same relation holds for $g_{\gamma\gamma}=7.75 \times 10^{-5}\,\gev$ with the values of all other parameters unchanged.

\subsection{Dark axion portal\label{sec:model_dark_alp}}

The interaction Lagrangian of the DAP is \cite{Kaneta:2016wvf,Ejlli:2016asd},
\be
\!\!\mathcal{L} & \supset \frac{g_{a\gamma\gamma^{\prime}}}{2} a F^{\mu\nu} \tilde{F'}_{\mu\nu}\,,
\label{eq:L_dark_axion}
\ee
where $g_{a\gamma\gamma^{\prime}}$ is a coupling of mass-dimension -1, $F_{\mu\nu}$ and $F'_{\mu\nu}$ are the EM and $U(1)_{\text{Dark}}$ field strength tensors, respectively, and the tilde symbol denotes the dual of the field strength tensor.

The dark axion portal leads to an interesting range of phenomena that are distinct from the photophilic ALP.
In particular, dark axion portal was proposed \cite{deNiverville:2018hrc} as an explanation of the recently rejuvenated $(g-2)_\mu$ anomaly \cite{Muong-2:2006rrc,Muong-2:2021ojo}.
The region of parameter space relevant to such a solution was the long-lived dark photon with $\sim\,$GeV mass.
In fact, \cite{deNiverville:2018hrc} analyzed the dark photon displaced decays in colliders, the past beam dump and neutrino experiments to exclude such possibility.
On the other hand, an extended dark axion portal, involving also kinetic mixing with SM hypercharge or muon-philic interactions, has been shown to be a viable solution \cite{Ge:2021cjz,Zhevlakov:2022vio}. 
Moreover, such a scenario could be tested in future lepton fixed target experiments, such as NA64$e$ \cite{Banerjee:2019pds}. These considerations further motivate dedicated sensitivity study of the long-lifetime regime of the DAP at the far-forward region of the LHC.

In the following, we discuss two benchmarks of the DAP described solely by \cref{eq:L_dark_axion}, where in each case one of the DS species is a massless, stable particle. 
Then only the coupling $g_{a\gamma\gamma^{\prime}}$ and the LLP mass are free parameters of the model.
In \cref{sec:results} we present results for both of these benchmarks, as well as for several additional scenarios in which the masses of the DS states follow a fixed ratio.

For both benchmarks, the lifetime of the unstable, and typically long-lived, particle depends on the width of the two-body decay into a photon and a DS state given by \cref{eq:Gamma_two_body}.
The three-body decays into a pair of charged leptons and a DS state are also possible, especially for $m\gtrsim 0.1\,\gev$, but they are phase-space suppressed. 
As a result, they will contribute to the total decay width typically only at the $O(0.01)$ level; relevant formulas are given by \cref{eq:gprime_lla,eq:a_llg}.

Since FASER detectors are $\sim 400-600\,\m$ away from the $p{\text-}p$ collision point of the LHC, the typical LLP decay lengths they can probe are 
\be
  d_{\gamma^\prime} \simeq &\, 100 \,\m \times \left(\frac{E}{1000\,\gev}\right) \left(\frac{0.1\,\gev}{m_{\gamma^\prime}}\right)^4 \left(\frac{7 \times 10^{-5}}{g_{a\gamma\gamma^\prime}}\right)^2,
  \label{eq:ctau_gprime}
\ee
for the massless dark axion, while for the massless dark photon analogous formula for $d_a$ holds for $g_{a\gamma\gamma^\prime}=4 \times 10^{-5}\,\gev^{-1}$.
We note that in the opposite mass hierarchy, $m_a \gg m_{\gamma^\prime}$, the lifetime of $a$ is smaller than the lifetime of $\gamma^\prime$ by a factor of $3$, coming from the average over dark photon polarization states. The same factor will occur for other pairs of processes in which $a$ and $\gamma^{\prime}$ are exchanged which will influence our results in \cref{sec:results}.

\subsection{Bino-ALPino\label{sec:bino_ALPino}}
The relevant part of the Lagrangian is \cite{Kim:1983ia,Kim:1984yn,Nieves:1986ed}
\be
  \!\!\mathcal{L} & \supset \frac{\alpha_{\mathrm{EM}} C_{a \gamma \gamma}}{16 \pi f_a} \overline{\tilde{a}} \gamma^5\left[\gamma^\mu, \gamma^\nu\right] \tilde{\chi}_0 F_{\mu \nu},
  \label{eq:L_chitilde_atilde}
\ee
where $\tilde{a}$ and $\tilde{\chi}_0$ denote the ALPino and neutralino fields, respectively, $\alpha_{\mathrm{EM}}$ is the fine structure constant, $C_{a \gamma \gamma}\sim O(1)$ is a mixing constant that depends on the ALP scenario \cite{Covi:1999ty,Covi:2001nw}, and $f_a$ denotes the PQ breaking scale.

The ALPino mass is in general a model-dependent quantity \cite{Chun:1992zk,Chun:1995hc,Choi:2013lwa}, so it essentially acts as a free parameter.
However, since the value of the ALPino mass will not significantly affect our discussion (as long as it is significantly smaller than the neutralino mass and does not cause large phase space suppression of the NLSP (next-to-lightest SUSY particle) decay width), we follow \cite{Choi:2019pos} and set its value as follows: $m_{\tilde{a}}=10\,\mev$.

In the case of a sub-GeV bino, the following benchmark corresponds to a sufficiently long-lived NLSP that can be probed by beam dump experiments:
\be
  d_{\tilde{\chi}_0} \simeq &\, 100 \,\m \times \left(\frac{E}{1000\,\gev}\right) \left(\frac{0.1\,\gev}{m_{\tilde{\chi}_0}}\right)^4 \left(\frac{f_a}{30\, \gev}\right)^2.
  \label{eq:ctau_gtilde_atilde}
\ee

The lifetime of a sub-GeV bino is determined by two-body decays given by \cref{eq:Gamma_axino_2body}, while three-body decays mediated by an off-shell photon typically contribute less than a percent, see \cref{eq:Gamma_axino_3body}.

\subsection{Bino-gravitino\label{sec:bino_gravitino}}
The interactions relevant for our study are described by the following Lagrangian \cite{Volkov:1972jx,Deser:1977uq,Wess:1992cp}:\footnote{To perform calculations, we follow the Feynman rules given in \cite{Pradler:2006tpx}.}
\be
  \!\!\mathcal{L} & \supset -\frac{i}{8 m_{\mathrm{Pl.}}} \bar{\psi}_\mu [\gamma^\rho, \gamma^\sigma] \gamma^\mu \tilde{\chi}_0 F_{\rho \sigma},
  \label{eq:L_chitilde_gtilde}
\ee
where $\psi_\mu$ denotes the gravitino wavefunction, and the Lorentz index indicates the spin-$3/2$ character of the field.

Compared to the ALPino model, mass of gravitino is not a free parameter.
Instead, the SUSY breaking energy scale determines it by the super-Higgs mechanism \cite{Volkov:1973jd,Deser:1977uq}. 
As a result, the gravitino mass is $m_{\tilde{G}} = F_{\mathrm{\mathrm{SUSY}}}/(\sqrt{3} \, m_{\mathrm{Pl.}})$.
Moreover, due to the SUSY Equivalence Theorem \cite{Casalbuoni:1988kv}, the gravitino wavefunction can be approximated at high energies as follows:\footnote{In our calculations, we instead take into the account all gravitino degrees of freedom through the gravitino polarization tensor given by \cref{eq:grav_tensor}.}
\be
  \psi_\mu \simeq i\sqrt{\frac 23} \frac{\partial_\mu \psi}{m_{\tilde{G}}},
\ee
where $\psi$ is the spin-$1/2$ goldstino absorbed by the gravitino. 
As a result, even though gravitino interactions are suppressed by the Planck mass (due to its character as a SUSY partner of the graviton), cf. \cref{eq:L_chitilde_gtilde}, the massive gravitino compensates this suppression by the $1/{m_{\tilde{G}}}$ factor.
Therefore, \textit{the bino-gravitino-photon coupling} is therefore proportional to the inverse of the square root of the SUSY breaking scale, $1/\sqrt{F_{\mathrm{SUSY}}}$, instead of being suppressed by the Planck mass.

For sub-GeV neutralinos, the long-lived regime corresponds to low-energy SUSY breaking scales,
\be
  d_{\tilde{\chi}_0} \simeq &\, 100 \,\m \times \left(\frac{E}{1000\,\gev}\right) \left(\frac{0.1\,\gev}{m_{\tilde{\chi}_0}}\right)^5 \left(\frac{F_{\mathrm{\mathrm{SUSY}}}}{(60 \, \gev)^2}\right)^2,
  \label{eq:ctau_gtilde_Gtilde}
\ee
where $d_{\tilde{\chi}_0}$ is the bino decay length in the laboratory reference frame.
Its lifetime is determined by decays into gravitino and photon, while decays into gravitino and $e^+ e^-$ pair are suppressed, cf. \cref{eq:Gamma_grav_2body,eq:Gamma_grav_3body}; see also the bottom panels of \cref{fig:results_gravitino}.

\begin{figure*}[tb]
  \centering
  \includegraphics[width=0.46\textwidth]{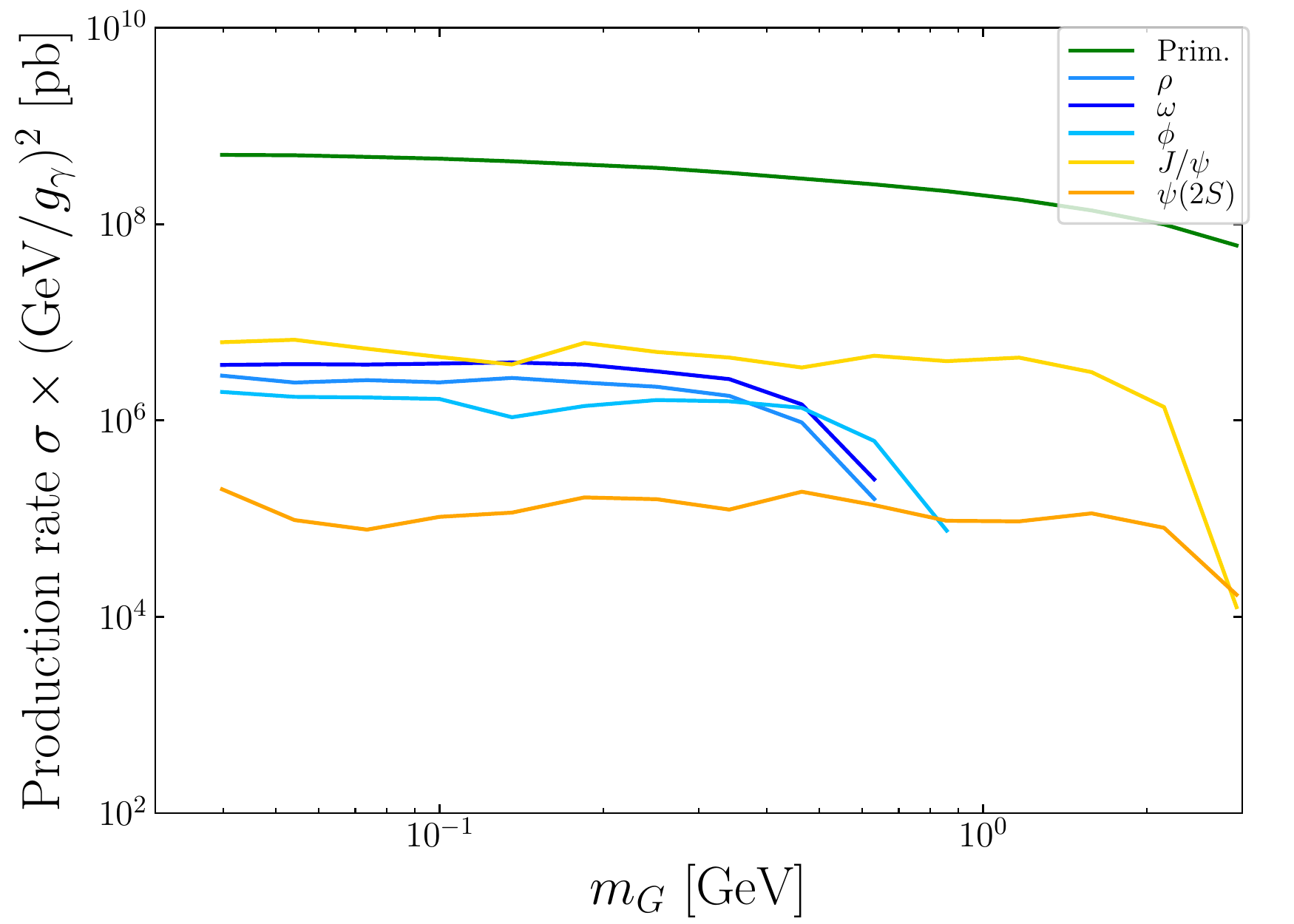}\hspace*{0.4cm}
  \includegraphics[width=0.46\textwidth]{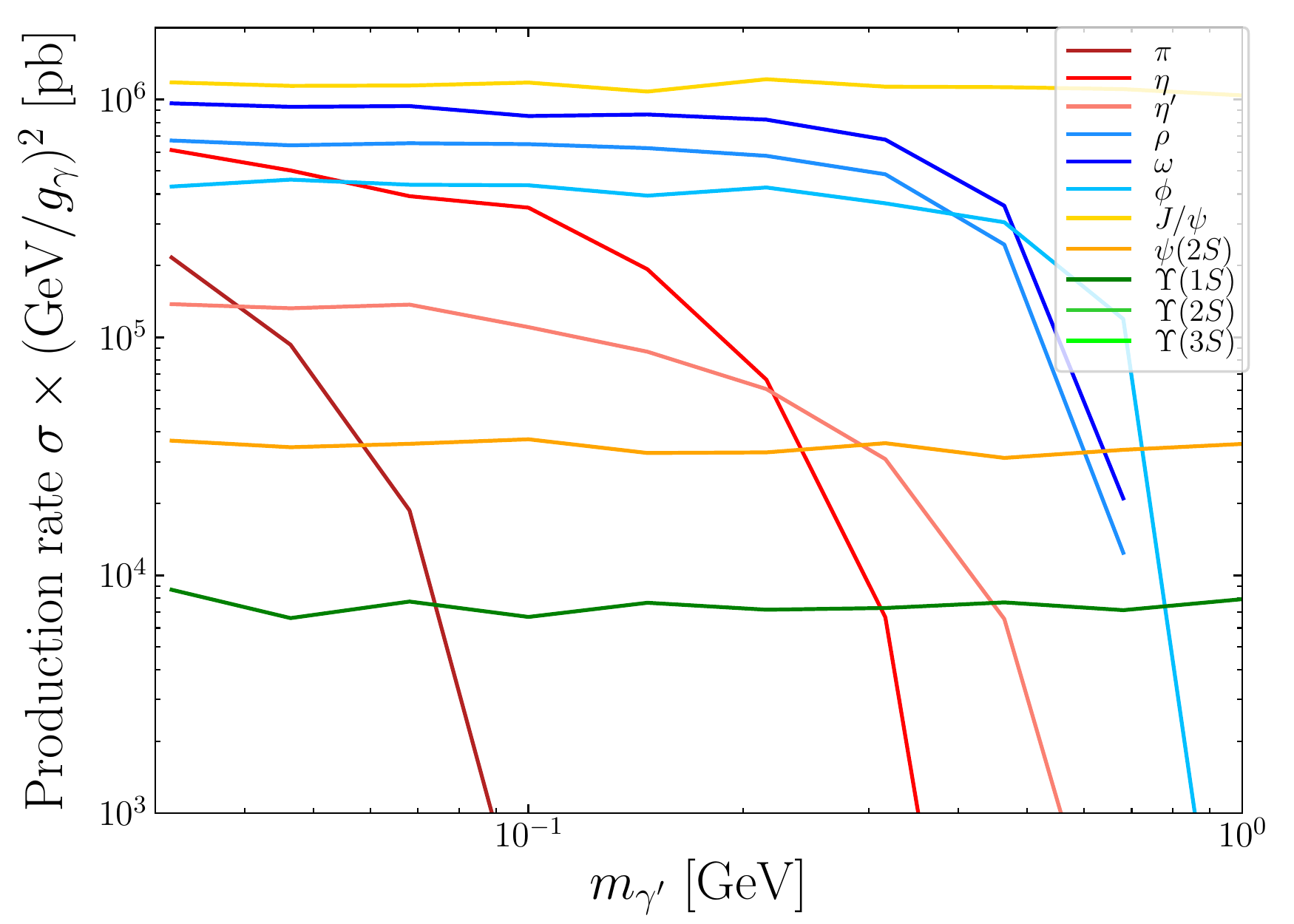}\vspace*{0.25cm}
  \includegraphics[width=0.46\textwidth]{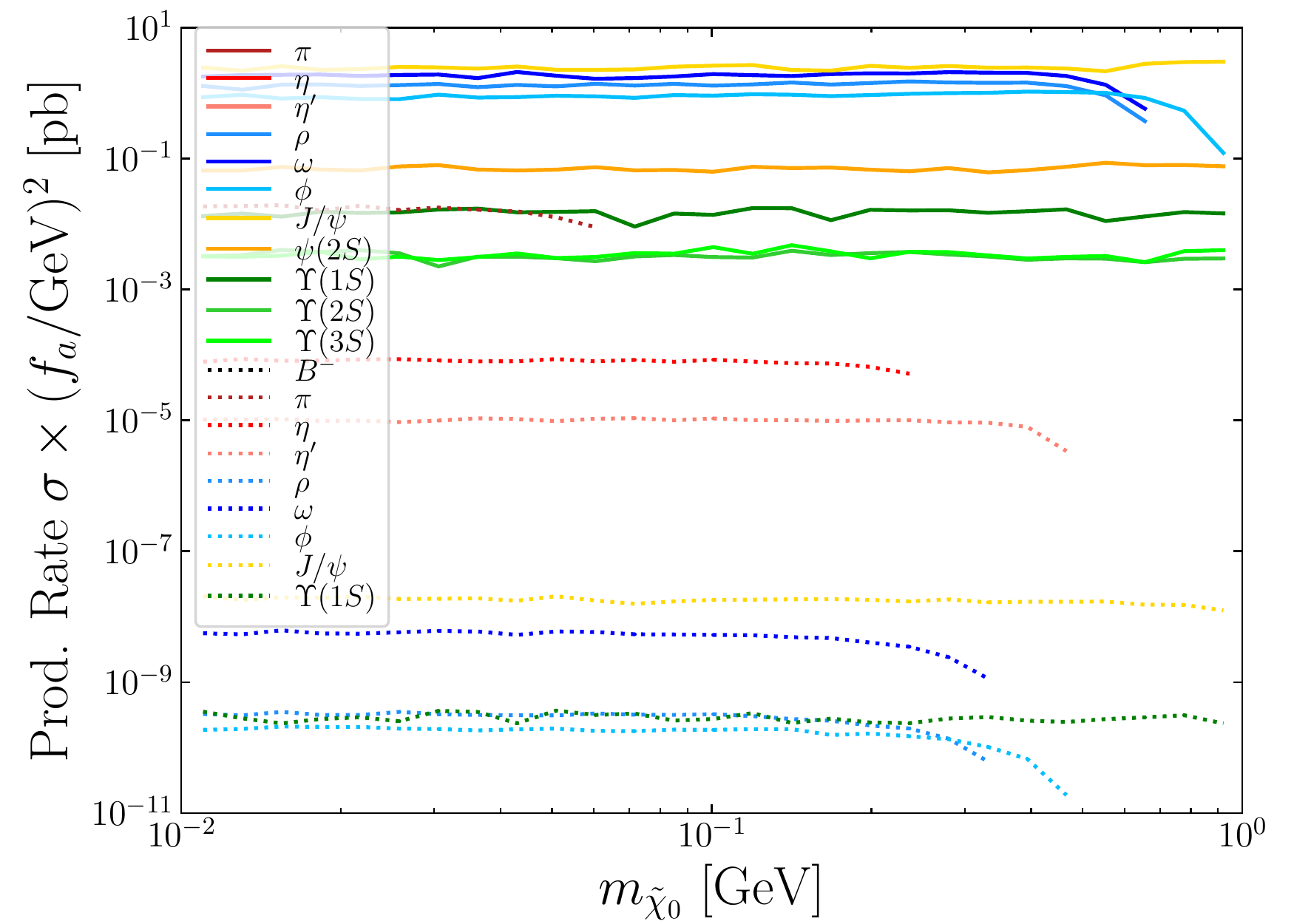}\hspace*{0.4cm}
  \includegraphics[width=0.46\textwidth]{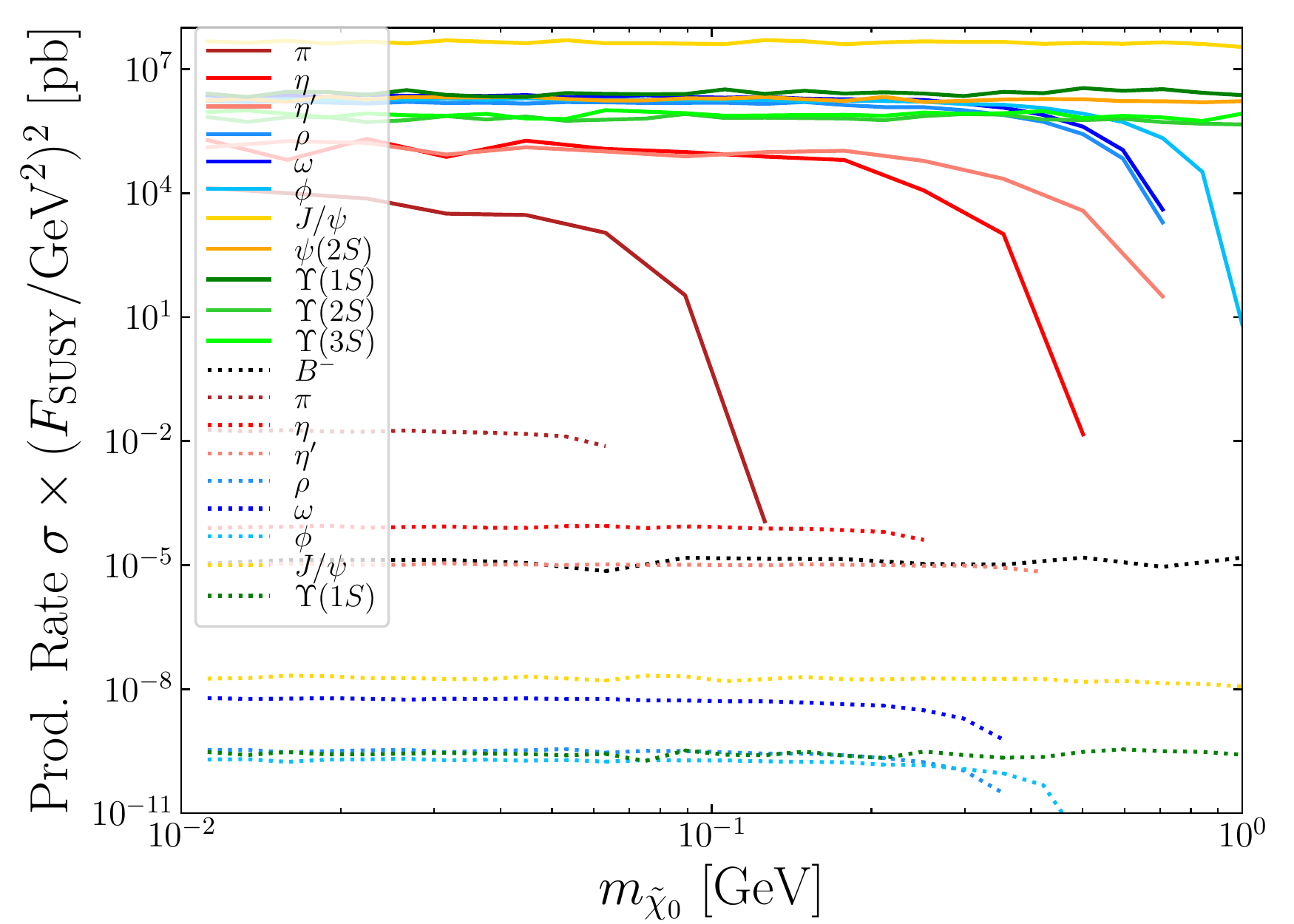}
  \caption{
    Yields of LLP production modes at FASER2 as a function of its mass; color coding indicates contributions of each mode according to the legend.
    \textit{Top left:} massive spin-2 mediator, where the direct production via Primakoff-like photon conversion (green) dominates over various vector meson decays.
    \textit{Top right:} dark photon acting as the LLP within dark axion portal, where the vector meson decays, which were not included in previous works, provide the leading contributions.
    \textit{Bottom:} Modes of light neutralino production for ALPino (\textit{left}) and gravitino (\textit{right}).
    The decays into neutralino-LSP depending on $f_a$ or $F_{\mathrm{SUSY}}$ are denoted by solid lines, while decays into a pair of neutralinos, which are independent of these couplings, are indicated by dotted lines. Such modes depend on the $m_{\mathrm{squark}}^{-4}$ instead.
    For ALPino, the $f_a$-independent decays dominate for the allowed values of $f_a$, while for gravitino, the decays depending on $F_{\mathrm{SUSY}}$ are the leading ones.
    }
  \label{fig:LLP_prod}
\end{figure*}

\section{LLP searches at the intensity frontier\label{sec:LLP_prod_and_sign}}
We investigate the prospects of detecting LLPs introduced in previous section in a number of upcoming experiments listed in \cref{tab:experiments}.
The top rows show the properties of (proper) beam dump experiments, while the bottom rows show the LHC-based detectors taking data in the far-forward direction.

\subsection{Monte Carlo simulation of LLPs}

\paragraph{LLP spectra}
For the forward direction detectors at the LHC, the $\tt FORESEE$ package can be used to obtain the LLP spectra. 
We extended it to also simulate the production and decay of LLPs taking place in beam dump experiments listed in \cref{tab:experiments}.
In this case, we used $\tt Pythia$ \cite{Sjostrand:2014zea} and $\tt BdNMC$ \cite{deNiverville:2016rqh} to generate unstable meson and photon spectra.
The photon spectrum generated by $\tt Pythia$ was validated with experimental data in \cite{Dobrich:2019dxc}, therefore, we use it in further analysis.

After the production of a LLP, the number of events linked to a LLP signature being detected inside the detector are \cite{Bauer:2018onh,Feng:2017uoz}
\be
  N = \int \int dE d\theta \frac{d^2 N}{dE d\theta}\, p(E, \theta)\, q_{\text{accept.}}(E, \theta),
  \label{eq:NoE}
\ee
where the first term denotes the spectrum of the LLP with a energy $E$ and polar angle $\theta$ relative to the beamline; $p(E)$ corresponds to the probability of the signature taking place inside the detector, while experimental or simulation-related cuts are described by $q_{\text{accept.}}(E, \theta, \phi)$.

\paragraph{Primary production}
Displaced LLP decays resulting from, \eg, proton-target collisions are the main experimental signature in LLP searches \cite{Battaglieri:2017aum,Beacham:2019nyx,Krnjaic:2022ozp}.
The experimental signal consists of high-energy SM particles, typically a pair of photons or charged leptons, and the probability of these decays occurring within a detector of length $\Delta$ is
\be
  p(E) = e^{-L/d(E)}-e^{-(L+\Delta)/d(E)},
  \label{eq:p_prim}
\ee
where $d(E)$ represents the LLP decay length in the LAB frame and $L$ corresponds to the distance between the LLP production point and the start of the detector. 
It is evident that the majority of events arise from sufficiently long-lived species, characterized by $d \gtrsim L$, resulting only in linear suppression with the decay length: $p(E) \simeq \Delta/d$ \cite{Essig:2013lka,Beacham:2019nyx}.
However, for short-lived species, the second term in \cref{eq:p_prim} can be neglected and $p(E) \simeq e^{-L/d}$. It is therefore clear that the distance $L$ sets the scale of the LLP decay lengths that can be probed in such a way.

For example, in the case of DAP, the leading two-body decays deposit energy through a single photon, while decays into a DS and $e^+ e^-$ are suppressed, see bottom panels of each plot in \cref{fig:results_dark_photon,fig:results_dark_axion}.
Despite the additional SM induced background for the single-photon LLP decay, it was shown \cite{Jodlowski:2020vhr} that FASER2 will be sensitive to it with the same cuts on the deposited energy and number of events as for the two-photon decays; we refer to that work for discussion of the backgrounds.

\paragraph{Secondary production}

Secondary production of LLPs can take place by coherent upscattering of a lighter DS species into the LLP on tungsten layers of neutrino emulsion detector FASER$\nu$2; see fig. 1 from \cite{Jodlowski:2019ycu} for a schematic illustration.

We study the displaced decay of the LLP produced in this way, where the production takes place at FASER$\nu$2, while the decay happens either at FASER2 or FASER$\nu$2. In the latter case, we demand the decay to take place at least $10\,\cm$ away from the upscattering point in order to avoid potential background from neutrino DIS.
As the distance between these two detectors is $L \simeq 1\,\m$, this production mode could allow to cover a part of the $d \sim L \simeq 1\,\m$ region of the parameter space.
On the other hand, the cross-section for the secondary production results in additional $\propto g^2$ dependence in the number of decays. 
As a result, the secondary LLP production can cover only larger values of $g$ than the ones covered by primary production.

The probability of secondary LLP production followed by decay inside FASER2 is given by the convolution of \cref{eq:p_prim} with upscattering cross-section, see \cite{Jodlowski:2019ycu} for discussion
\begin{widetext}
  \be
    p(E)_{\text{sec.\! prod.}} = \frac{1}{L_{\text{int}}} \int_{0}^{\tilde{\Delta}} \left(e^{-(L-t)/d}-e^{-(L+\Delta-t)/d}\right)\, dt = \frac{d}{m_T/(\rho\, \sigma(E))} e^{-(L+\Delta)/d} \left(e^{\Delta/d}-1\right) \left(e^{\tilde{\Delta}/d}-1\right),
    \label{eq:p_sec}
  \ee
\end{widetext}
where $L_{\text{int}}=m_T/(\rho\, \sigma(E))$ is the interaction length corresponding to the upscattering of DS species with energy $E$ on nucleus of mass $m_T$ inside the material of density $\rho$ and length $\tilde{\Delta}$; $\sigma(E)$ is the upscattering cross-section; $L$ is the distance from the beginning of the upscattering material to the beginning of the detector of length $\Delta$; and the dummy variable $t$ parameterizes the length of the upscattering material.

\paragraph{Electron scattering}
FASER$\nu$2 and FLArE detectors will be also sensitive to DS states scattering with electrons, see \cite{Batell:2021blf} for an extensive discussion; we follow the experimental cuts on electron scattering signature proposed in this study.

The corresponding probability for such scattering events is simply given by
\be
  p(E)_{\text{scat.}} = \frac{\Delta}{L_{\text{int}}},
  \label{eq:p_sec}
\ee
where $\Delta$ is the length of the FASER$\nu$2 or FLArE and $L_{\text{int}}$ denotes the interaction length of the scattering process.

\subsection{Experiments}
\paragraph{Beam dumps}
Beam dump experiments employ a beam of high-energy, $O(10-100)\,\gev$, protons striking a target composed of typically dense nuclei.
This results in a hadronic cascade, producing many unstable SM particles.
Although the decays of such states are generally well understood \cite{Workman:2022ynf}, the luminosity of the beam is usually high enough that even strongly suppressed branching ratios of decays into DS states can result in a sizable production of BSM LLPs.

We study $G$ decays into a pair of photons or charged leptons in past detectors such as CHARM \cite{CHARM:1985anb}, NuCal \cite{Blumlein:1990ay,Blumlein:2011mv}, as well as in the future detectors such as NA62 \cite{Dobrich:2018ezn}, SeaQuest \cite{Berlin:2018pwi}, and SHiP \cite{SHiP:2015vad,Alekhin:2015byh}.
Although their modus operandi is similar, they differ not only by size, geometry or the beam luminosity, but also in energy and target for high-energy protons, as well as different energy thresholds for the energy deposited by LLP decays.
As a result, they probe diverse LLP setups.

It is worth pointing out that although beam dump experiments have been in use for many decades, see recent review \cite{Lanfranchi:2020crw} for an overview of their past and recent results, LLP searches at colliders, mainly the LHC, are developing intensively, see, \eg, \cite{Knapen:2022afb}.
In the next subsection, we describe one such approach using the far-forward detectors at the LHC.

\paragraph{Forward direction detectors at the LHC}
The LHC can be used as an abundant source of high-energy photons nearly collimated along the direction of the proton-proton collisions \cite{Feng:2018pew}. After travelling $\sim 100\, m$, these photons, and other neutral particle, are absorbed by a thick block of iron called the TAN, effectively acting as a fixed-target beam dump. 
A fraction of the incoming photons are converted into ALPs by the Primakoff \cite{Primakoff:1951iae,Tsai:1986tx} coherent upscattering on the TAN nuclei.
As a result, even a small detector, FASER \cite{Feng:2017uoz,Feng:2017vli}, located at a considerable distance from the photon production site can search for an ALP decaying into two photons in an essentially background-free manner.\footnote{Searches for other LLPs decaying typically into the SM charged leptons can be performed in a similar way \cite{FASER:2018eoc}.}
FASER has been collecting data since 2022 \cite{FASER:2022hcn} and, in addition to the main detector, includes a neutrino emulsion detector FASER$\nu$ \cite{FASER:2019dxq,FASER:2020gpr}, which is located upstream of the main detector.
Further research in this direction has resulted in a number of proposals for significant extensions to the original FASER experiment, for example, FASER2 \cite{FASER:2018ceo,FASER:2018bac,FASER:2021ljd}, or an entirely separate facility called the Forward Physics Facility (FPF) \cite{MammenAbraham:2020hex,Anchordoqui:2021ghd,Feng:2022inv}, which would house a number of detectors dedicated to various complementary searches. Among these are: AdvSND \cite{Boyarsky:2021moj}, FASER$\nu$2 \cite{Batell:2021blf,Anchordoqui:2021ghd}, FLArE \cite{Batell:2021blf}, and FORMOSA \cite{Foroughi-Abari:2020qar}.\footnote{We note there are many more LLP detectors proposals which are at various stages of progress, see \cite{Ilten:2022lfq} for extensive overview. Among them are, \eg, Codex-B \cite{Aielli:2019ivi}, FACET \cite{Cerci:2021nlb}, and milliQan \cite{milliQan:2021lne}.}

Another aspect that makes FASER2/FPF particularly well-positioned for studying photon-coupled DS species is the presence of the tungsten neutrino detector FASER$\nu$2 \cite{FASER:2019dxq,FASER:2020gpr} located in front of the main detector (decay vessel). 
As our previous work has shown \cite{Jodlowski:2019ycu,Jodlowski:2020vhr}, a secondary production of LLPs can take place at FASER$\nu$2 through coherent upscattering of stable DS species on tungsten nuclei which allows to probe the shorter LLP lifetime regime. We also explore this further in other work \cite{Jodlowski:2023ohn}.

FASER$\nu$2 and FLArE detectors \cite{Batell:2021blf} will also be able to probe BSM scenarios by scattering of DS species with electrons or by converting a LLP into a single, high-energy photon \cite{Kling:2022ykt,Kling:2022ehv}.
In the spin-2 portal the latter signature is particularly effective because it occurs through coherent scattering with a nucleus, which is enhanced by a $Z^2$ factor, and takes place by a photon exchange whose propagator, $1/t$, is enhanced in such a low-momentum exchange process.
The corresponding probability for $G\to \gamma$ conversion is
\be
  p(E)_{\text{scat.}} = \frac{\Delta}{L_{\text{int}}},
  \label{eq:p_sec}
\ee
where $L_{\text{int}}=m_T/(\rho\, \sigma(E))$ denotes the interaction length of the conversion process whose cross-section is $\sigma(E)$, $m_T$ is the mass of the nucleus inside the material of density $\rho$ and length $\Delta$.

\begin{table*}[h]
  \centering
  \hspace*{-1cm}
  \begin{tabular}{|c||c|c|c|c|c|c|c|c|c|}
    \hline
    \hline
    Experiment & \thead{Target for \\ prim/sec. \\ prod.} & Energy & \thead{Lumi.\\or $N_{\mathrm{prot.}}$} & \thead{Transverse \\ size} & $L$ & $\Delta$ & LLP signature & \thead{LLP \\signature cuts} & Ref. \\
    \hline
    \hline
    CHARM & Cu/- & 400 GeV & $2.4 \times 10^{18}$ & $3\times 3$ m$^2$\footnote{The detector was placed 5 m away from the beam axis; other detectors are on-axis.} & 480 m & 35 m & decay & \thead{$E_{e^+ e^-}>3\ \gev$: $N_{\mathrm{ev}}=3$ \\ $E_{\gamma \gamma}>3\ \gev$: $N_{\mathrm{ev}}=3$ \\ $E_{\gamma}>7.5\ \gev$: $N_{\mathrm{ev}}=100$} & \cite{Blumlein:2013cua,Dobrich:2019dxc} \\
    \hline
    NA62 & Cu/- & 400 GeV & $1.0 \times 10^{18}$ & $r=1.13$ m & 81 m & 135 m & decay & \thead{$E_{e^+ e^-}>3\ \gev$: $N_{\mathrm{ev}}=3$ \\ $E_{\gamma\gamma}>3\ \gev$: $N_{\mathrm{ev}}=3$}  & \cite{Dobrich:2019dxc} \\
    \hline
    NuCal & Fe/- & 69 GeV & $1.7 \times 10^{18}$ & $r=1.3$ m & 23 m & 64 m & decay & \thead{$E_{e^+ e^-}>10\ \gev$: $N_{\mathrm{ev}}=4.4$ \\ $E_{\gamma\gamma}>10\ \gev$: $N_{\mathrm{ev}}=4.4$ \\ $E_{\gamma}>10\ \gev$: $N_{\mathrm{ev}}=4.4$} & \cite{Blumlein:2013cua,Dobrich:2019dxc} \\
    \hline
    SeaQuest & Fe/- & 120 GeV & $1.44 \times 10^{18}$ & $2\times 2$ m$^2$ & 5 m & 0.95 m & decay & \thead{$E_{e^+ e^-}>3\ \gev$: $N_{\mathrm{ev}}=3$ \\ $E_{\gamma\gamma}>3\ \gev$: $N_{\mathrm{ev}}=3$} & \cite{Dobrich:2019dxc,Choi:2019pos,Blinov:2021say} \\
    \hline
    SHiP & Mo/- & 400 GeV & $2.4 \times 10^{18}$ & $2.5\times 5.5$ m$^2$ & 52.7 m & 50 m & decay & \thead{$E_{e^+ e^-}>3\ \gev$: $N_{\mathrm{ev}}=3$ \\ $E_{\gamma\gamma}>3\ \gev$: $N_{\mathrm{ev}}=3$ \\ $E_{\gamma}>2\ \gev$: $N_{\mathrm{ev}}=100$}  & \cite{Dobrich:2019dxc,Jodlowski:2019ycu} \\
    \hline
    \hline
    FASER2 & Fe\footnote{By primary LLP production at the LHC, we mean the Primakoff process in which photons produced in pp collisions hit the iron hadronic absorber TAN located 140 m further converting into a LLP particle; the same is assumed for other versions of FASER detector.}/- & \thead{$\sqrt{s}=$ \\ $13\,\tev$} & $3000$ fb$^{-1}$ & $r=1$ m & 480 m & 5 m & decay  & \thead{$E_{e^+ e^-}>100\ \gev$: $N_{\mathrm{ev}}=3$ \\ $E_{\gamma\gamma}>100\ \gev$: $N_{\mathrm{ev}}=3$ \\ $E_{\gamma}>100\ \gev$: $N_{\mathrm{ev}}=3$}  & \cite{Feng:2018pew,Jodlowski:2019ycu} \\
    \hline
    FASER$\nu$2 & Fe/W & \thead{$\sqrt{s}=$ \\ $13\,\tev$} & $3000$ fb$^{-1}$ & $r=0.25$ m & 472 m & 2 m & \thead{conv. into $\gamma$ \\ decay, \\ sec. prod., \\ $e^-$ scat.} & \thead{$E^{\mathrm{conv.}}_{G \to \gamma}>1000\ \gev$: \\ $N_{\mathrm{ev}}=3$ \\ $E_{\gamma}>1000\ \gev$: $N_{\mathrm{ev}}=3$, \\ $300\, \mev<E_{e^-}< 20\, \gev$:\footnote{For FASER$\nu$2 and FLArE we also take into the account the angular cuts - see tables 1 and 2 from \cite{Batell:2021blf}.} \\ $N_{\text{ev}}=20$} & \cite{Jodlowski:2020vhr,Kling:2022ehv} \\
    \hline
    FPF FASER2 & Fe/W & \thead{$\sqrt{s}=$ \\ $13\,\tev$} & $3000$ fb$^{-1}$ & $r=1$ m & 620 m & 25 m & decay  & \thead{$E_{e^+ e^-}>100\ \gev$: $N_{\mathrm{ev}}=3$ \\ $E_{\gamma\gamma}>100\ \gev$: $N_{\mathrm{ev}}=3$ \\ $E_{\gamma}>100\ \gev$: $N_{\mathrm{ev}}=3$} & \cite{Feng:2018pew,Feng:2022inv,Jodlowski:2020vhr} \\
    \hline
    FPF FASER$\nu$2 & Fe/W & \thead{$\sqrt{s}=$ \\ $13\,\tev$} & $3000$ fb$^{-1}$ & $0.4\times 0.4$ m$^2$ & 612 m & 8 m & \thead{conv. into $\gamma$ \\ decay, \\ sec. prod., \\ $e^-$ scat.} & \thead{$E^{\mathrm{conv.}}_{G \to \gamma}>1000\ \gev$: \\ $N_{\mathrm{ev}}=3$ \\ $E_{\gamma}>1000\ \gev$: $N_{\mathrm{ev}}=3$, \\ $300\, \mev<E_{e^-}< 20\, \gev$: \\ $N_{\text{ev}}=20$} & \cite{Jodlowski:2019ycu,Feng:2022inv,Jodlowski:2020vhr} \\
    \hline
    FPF FLArE & Fe/Ar & \thead{$\sqrt{s}=$ \\ $13\,\tev$} & $3000$ fb$^{-1}$ & $1\times 1$ m$^2$ & 600 m &  7 m & \thead{conv. into $\gamma$ \\ sec. prod., \\ $e^-$ scat.}  & \thead{$E^{\mathrm{conv.}}_{G \to \gamma}>1\ \gev$: \\ $N_{\mathrm{ev}}=3$ \\ $30\, \mev<E_{e^-}< 1\, \gev$: \\ $N_{\mathrm{ev}}=20$} & \cite{Batell:2021blf,Kling:2022ykt,Kling:2022ehv,Feng:2022inv} \\
    \hline
    MATHUSLA & -/Si & \thead{$\sqrt{s}=$ \\ $13\,\tev$} & $3000$ fb$^{-1}$ &  -\footnote{MATHUSLA is proposed to be placed highly off-axis from the LHC pp beam; see Fig. 4 from \cite{Jodlowski:2019ycu} for illustration of its geometry.} & -  &  -  & \thead{decay} & \thead{$E_{e^+ e^-}>2\ \gev$: $N_{\mathrm{ev}}=3$} & \cite{Jodlowski:2019ycu} \\
    \hline
    \hline
  \end{tabular}
  \caption{
    Specification of considered detectors sensitive to LLPs decays or other signatures. We provide the technical parameters for each of the experiments used in our simulation, with references where details of each detector, and cuts on the LLP signatures, are given.
    Experiments utilizing the LHC $p{\text-}p$ beam are separated from beam dumps.
  }
  \label{tab:experiments}
\end{table*}

\clearpage

\begin{figure*}[tb]
  \centering
  \includegraphics[width=0.48\textwidth]{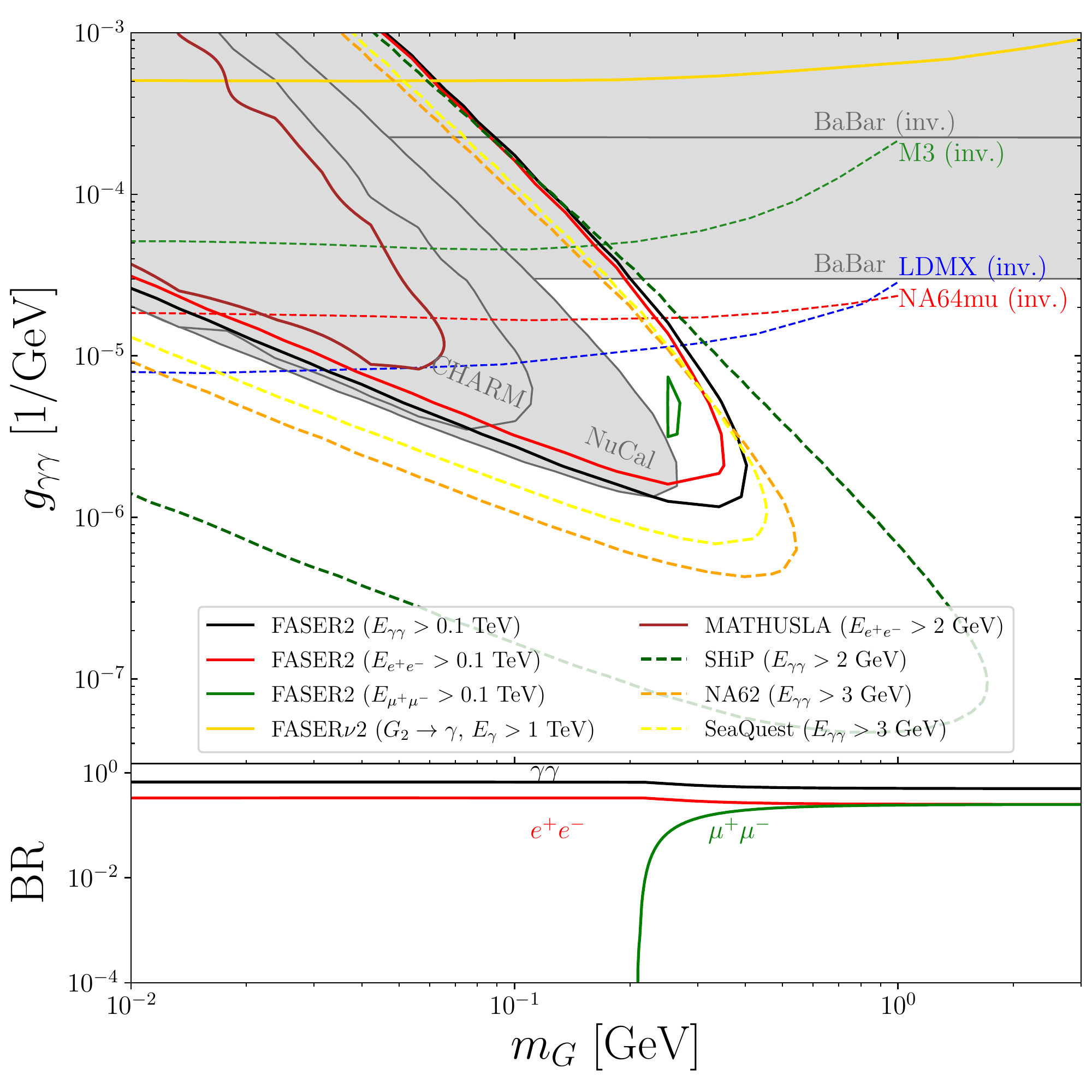} \hspace*{0.4cm} \includegraphics[width=0.48\textwidth]{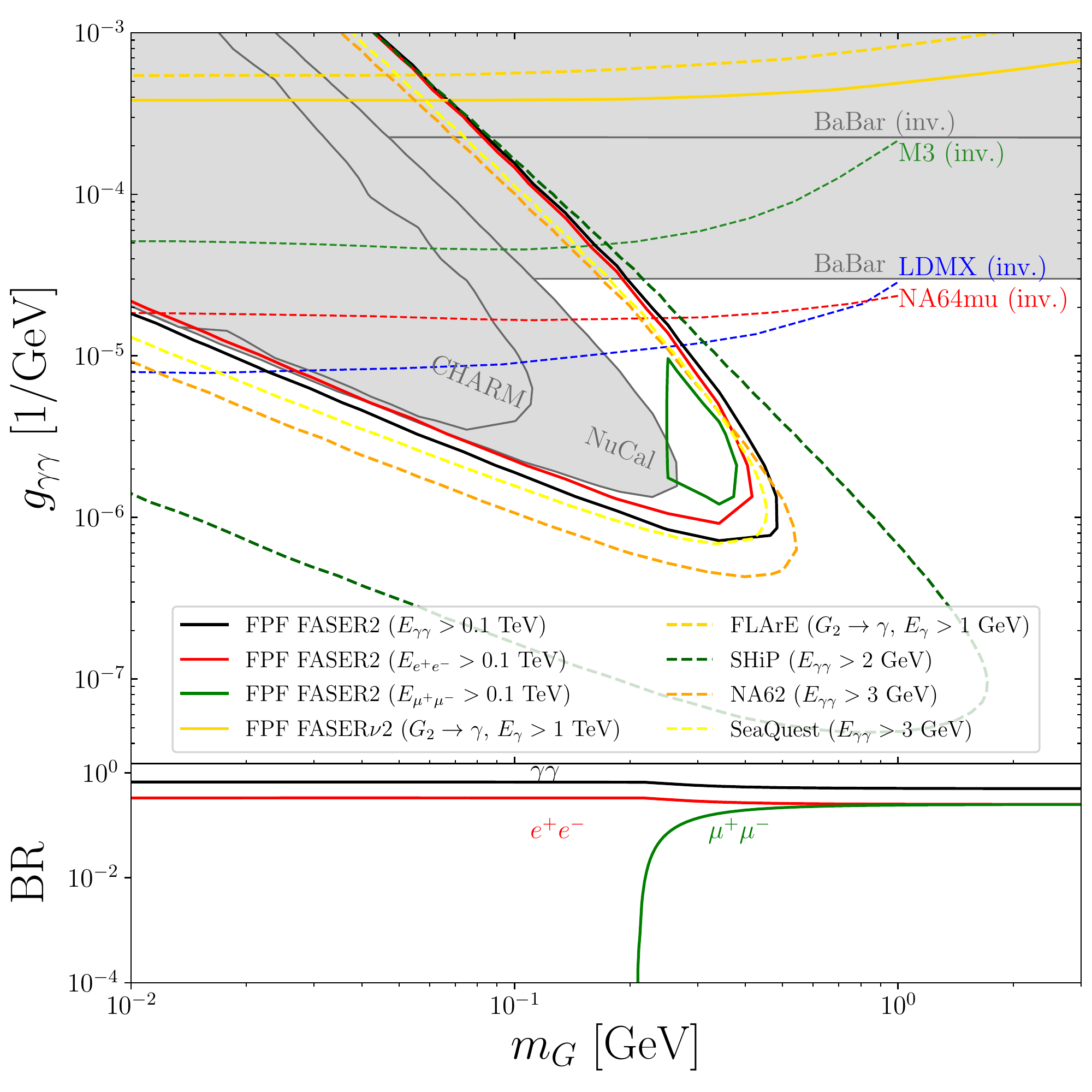}
  \caption{Sensitivity reaches for the universal coupling of $G$ to the SM gauge and matter fields. 
  We consider two setups of the FASER2 experiment: the baseline scenario \textit{(left)} and the Forward Physics Facility containing additional detectors, \eg, FLArE \textit{(right)}.
  We plot the contour lines of the number of events, $N_{\mathrm{ev}}$, for each of the experiments indicated in \cref{tab:experiments}, while gray-shaded regions denote current exclusion bounds from BaBar, CHARM, and NuCal. 
  Colorful lines not indicated in the plot legend were obtained in \cite{Voronchikhin:2022rwc} by considering the missing energy signature.
  }
  \label{fig:results_univ}
\end{figure*}

\begin{figure*}[tb]
  \centering
  \includegraphics[width=0.48\textwidth]{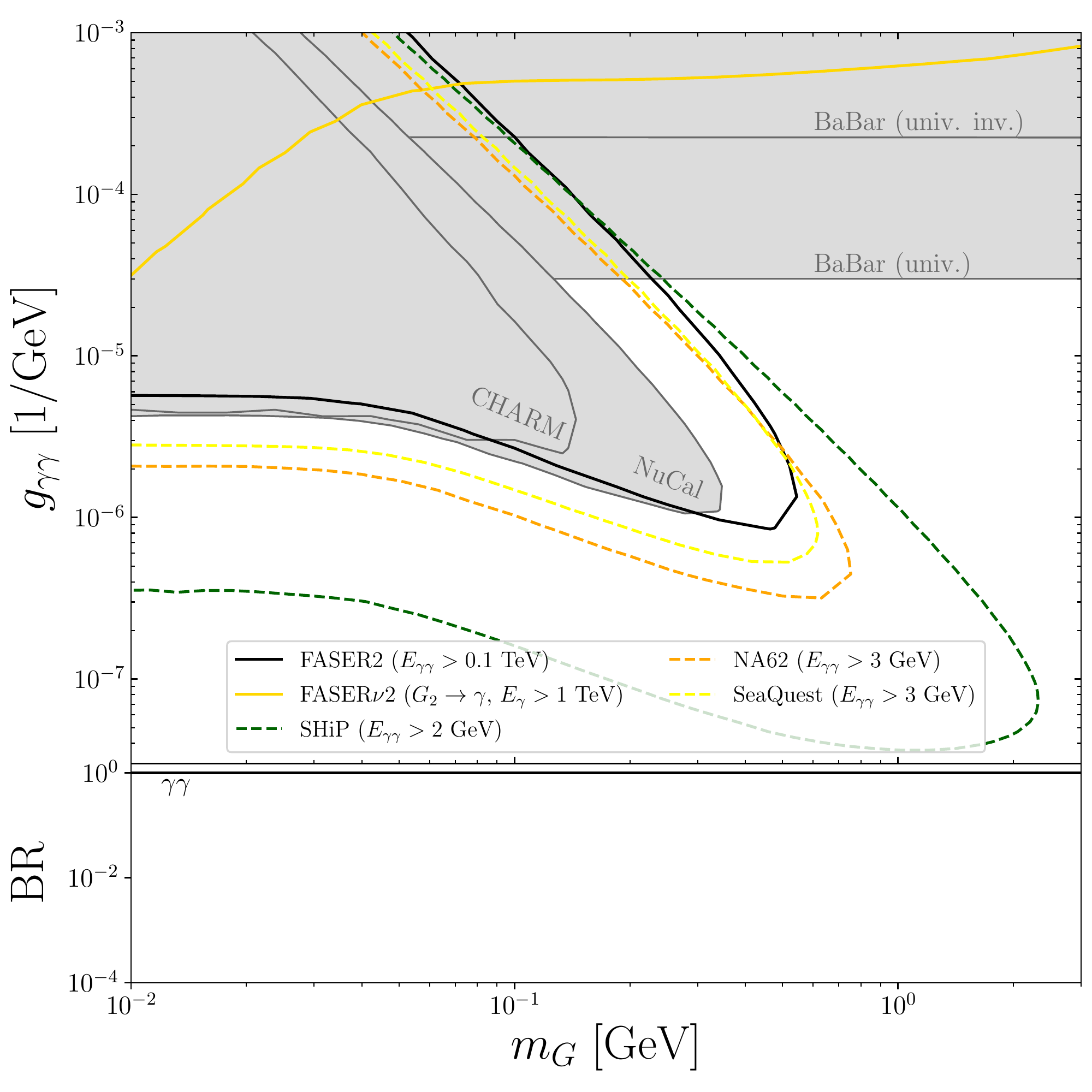} \hspace*{0.4cm} \includegraphics[width=0.48\textwidth]{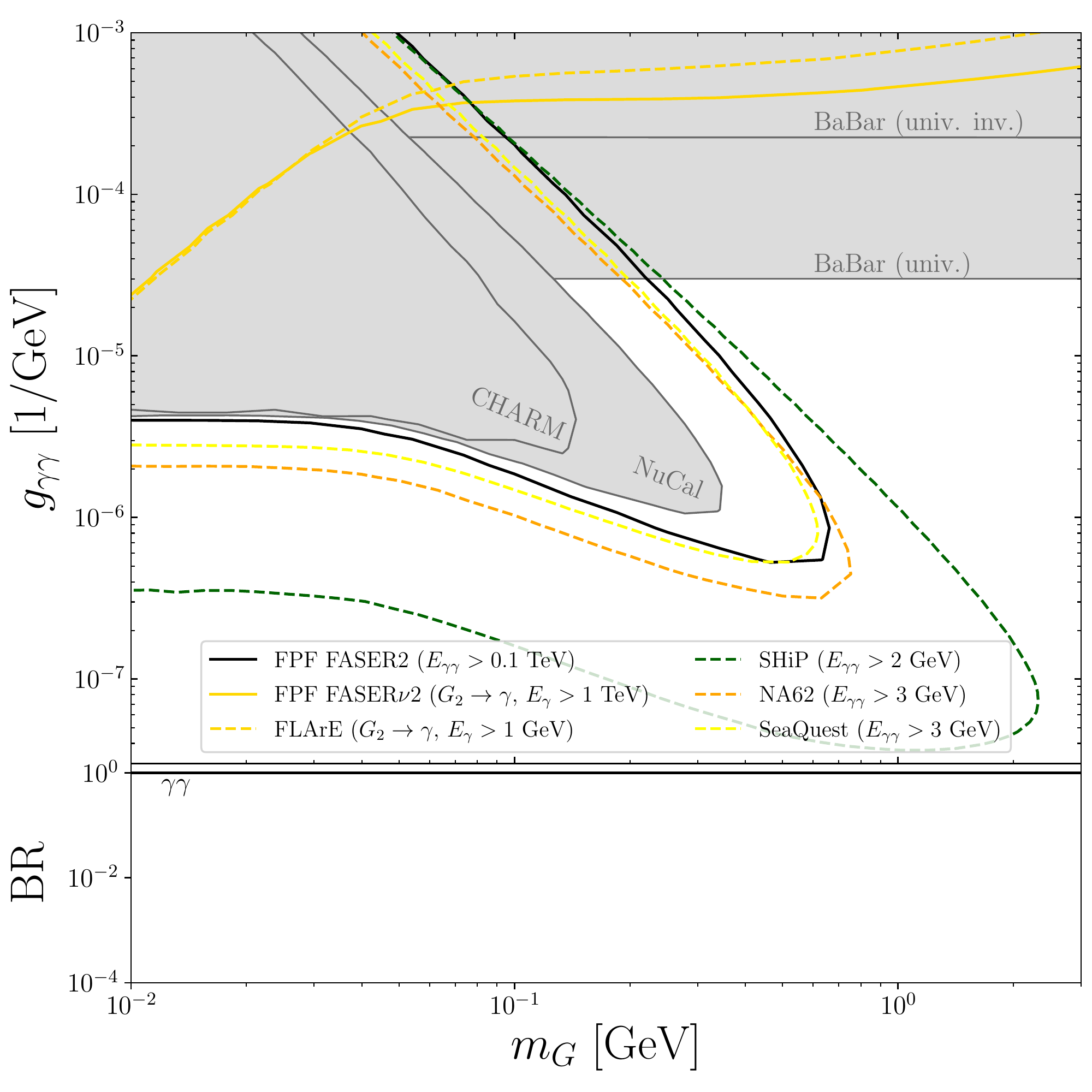}
  \caption{Same as \cref{fig:results_univ}, but for the model with non-universal, photophilic coupling. The saturation of the LLP decay sensitivity lines for $m_G \lesssim 50\,\mev$ is caused by the loss of perturbative unitarity for the dominant $G$ production mode, the Primakoff process, which depends on the $G$ mass as $\sigma \propto 1/m_G^4$.
  This is canceled by the factor $1/d \propto m_G^4$, which comes from the decay probability in the formula describing the total number of events, \cref{eq:NoE}.
  }
  \label{fig:results_nonuniversal}
\end{figure*}

\begin{figure*}[tb]
  \centering
  \includegraphics[width=0.48\textwidth]{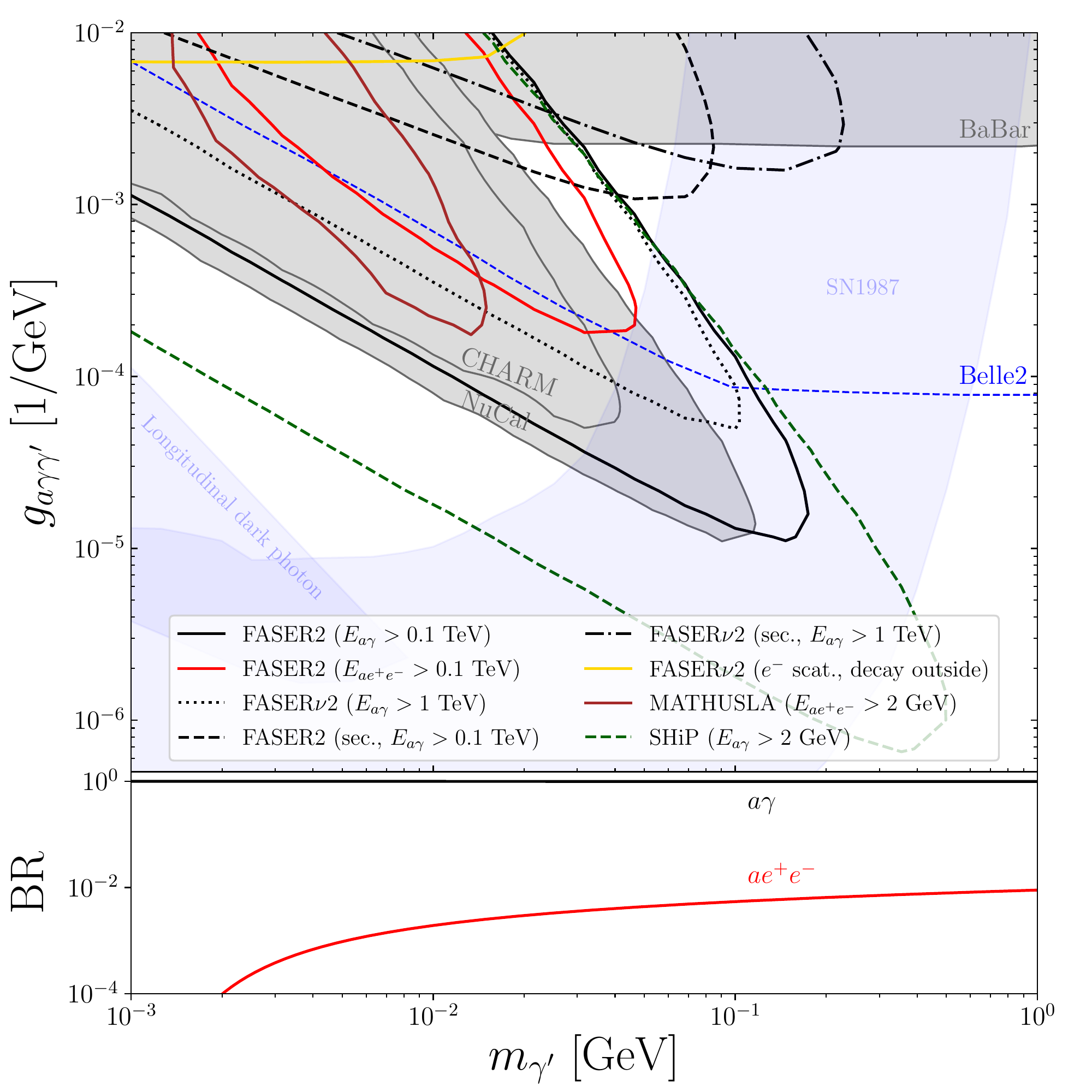}
  \hspace*{0.4cm}
  \includegraphics[width=0.48\textwidth]{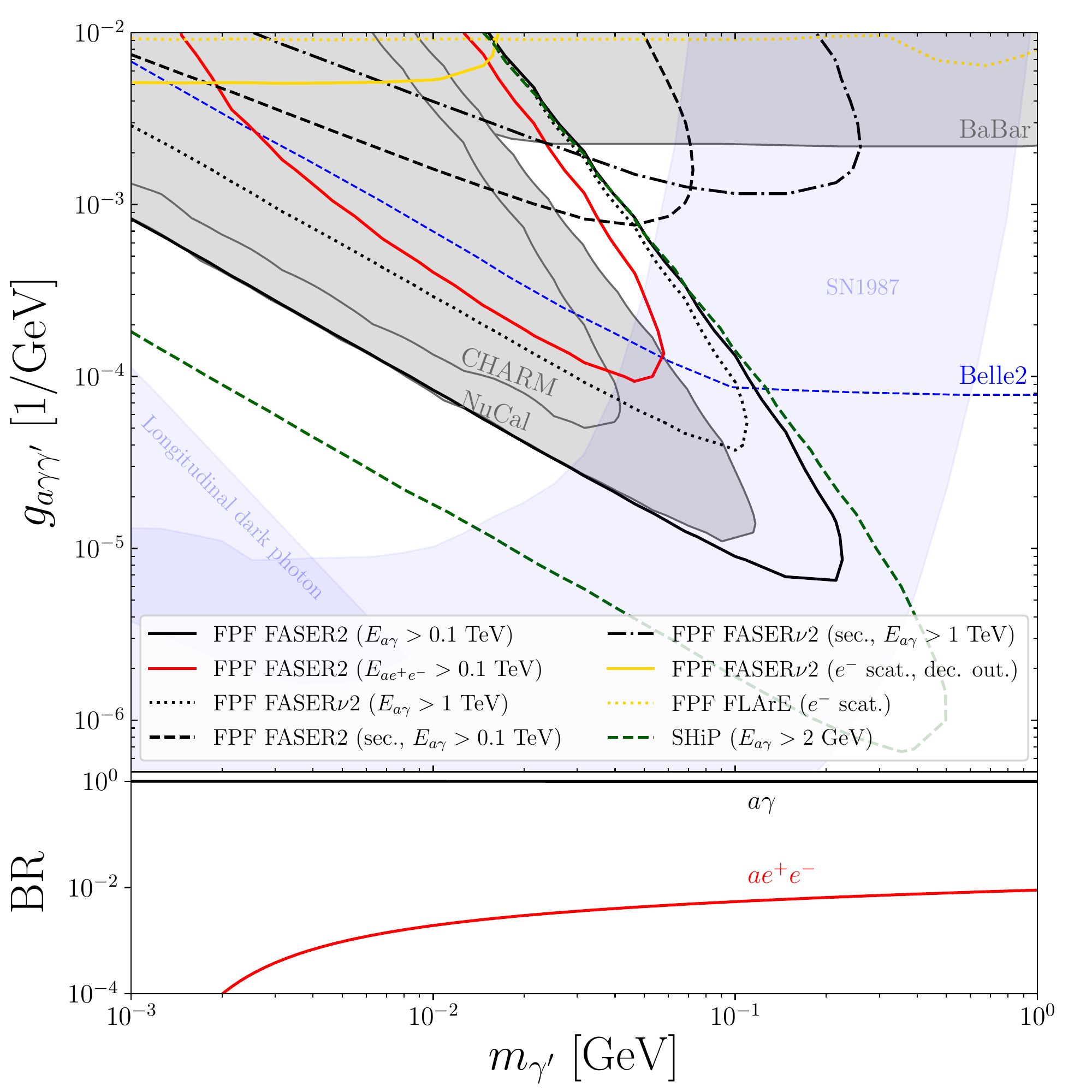}
  \caption{
    Sensitivity reach for the dark photon acting as the LLP while the dark axion is massless at the baseline (left) and the Forward Physics Facility (right) location of FASER2.
    The contour lines for each experiment correspond to the number of events, $N_{\mathrm{ev}}$, as indicated in \cref{tab:experiments}.
    Lines derived by the missing energy signature at BaBar and Belle were taken from \cite{deNiverville:2018hrc}.
    We also present the dark photon decay branching ratios - it shows that the ability to detect a single high-energy photon significantly extends the reach of FASER2.
    Moreover, the larger values of mass and coupling constant can be partially covered thanks to secondary LLP production. We also indicate the visible energy thresholds used to simulate the dark photon decays.
    }
  \label{fig:results_dark_photon}
\end{figure*}

\begin{figure*}[tb]
  \centering
  \includegraphics[width=0.95\textwidth]{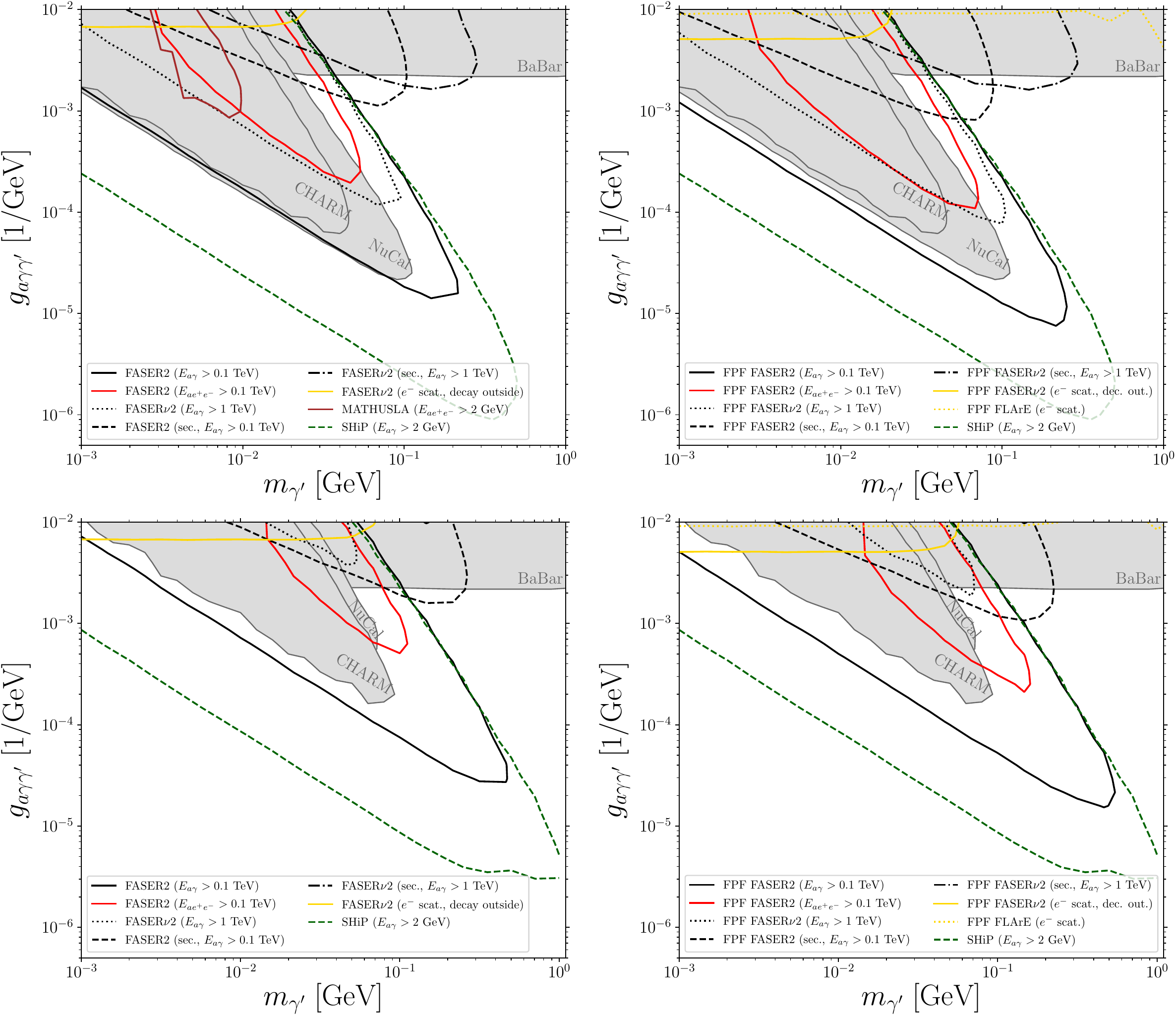}
  \caption{
    Same as \cref{fig:results_dark_photon}, but for the mass ratio, $m_{a}/m_{\gamma^\prime}$, fixed as follows: $0.5$ (top) and $0.9$ (bottom).
    As a result, the LLP lifetime is extended, shifting the sensitivity reach of each experiment toward larger masses. 
    In addition, as is particularly evident in the case of the $0.9$ mass ratio, the visible energy in $\gamma^\prime$ decays is phase-space suppressed.
    As a result, the bounds are considerably relaxed, especially in the case of NuCal. However, since secondary LLP production requires very energetic LLPs, $E>100\,\gev$, this effect does not affect the main signature, which is secondary LLP production followed by decay inside FASER2 (dashed line), except for decays inside FASER$\nu2$ (dashed-dotted line), since the latter requires $E_\gamma>1\,\tev$ instead.
    }
  \label{fig:results_dark_photon_mass_ratio_0.5_0.9}
\end{figure*}

\begin{figure*}[tb]
  \centering
  \includegraphics[width=0.48\textwidth]{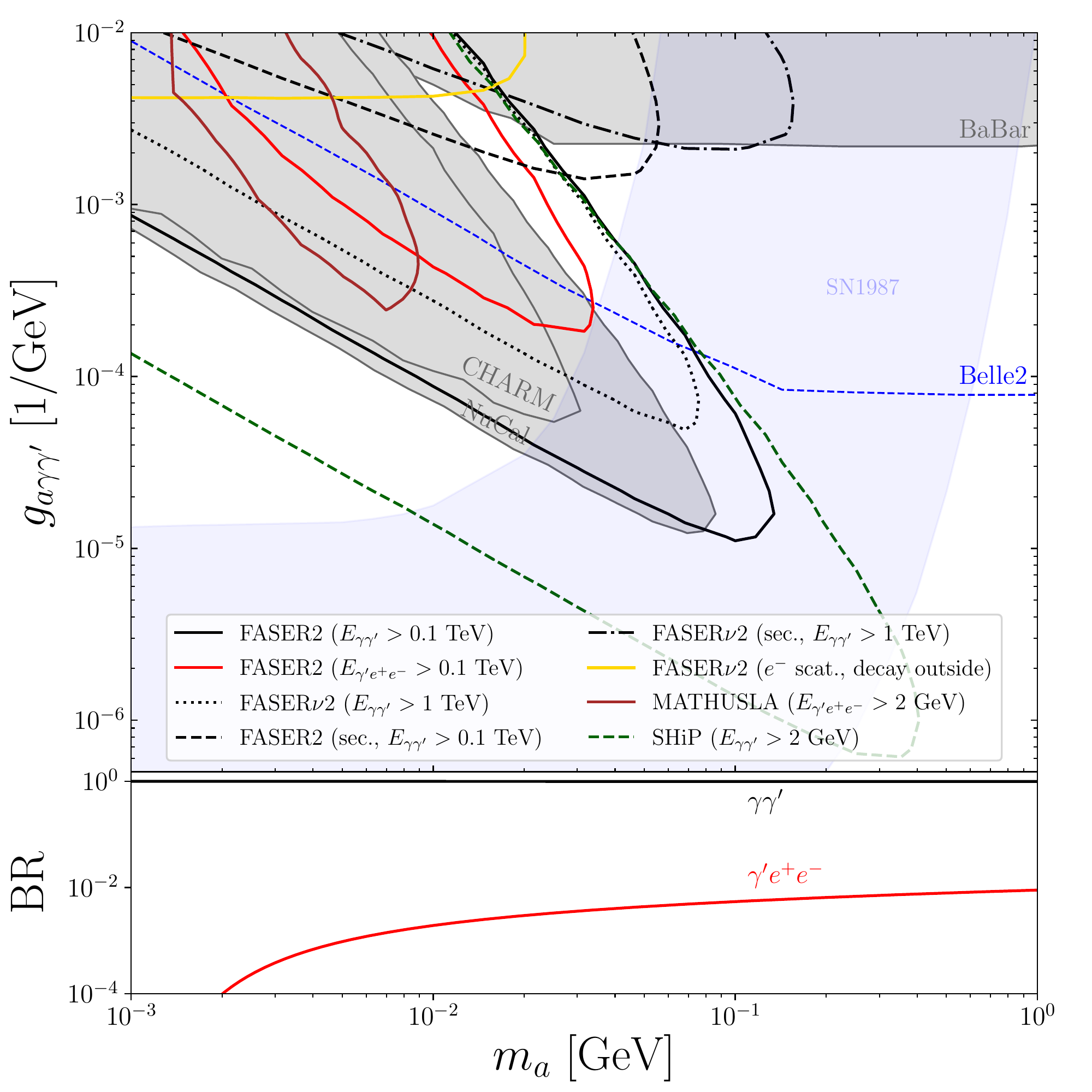}
  \hspace*{0.4cm}
  \includegraphics[width=0.48\textwidth]{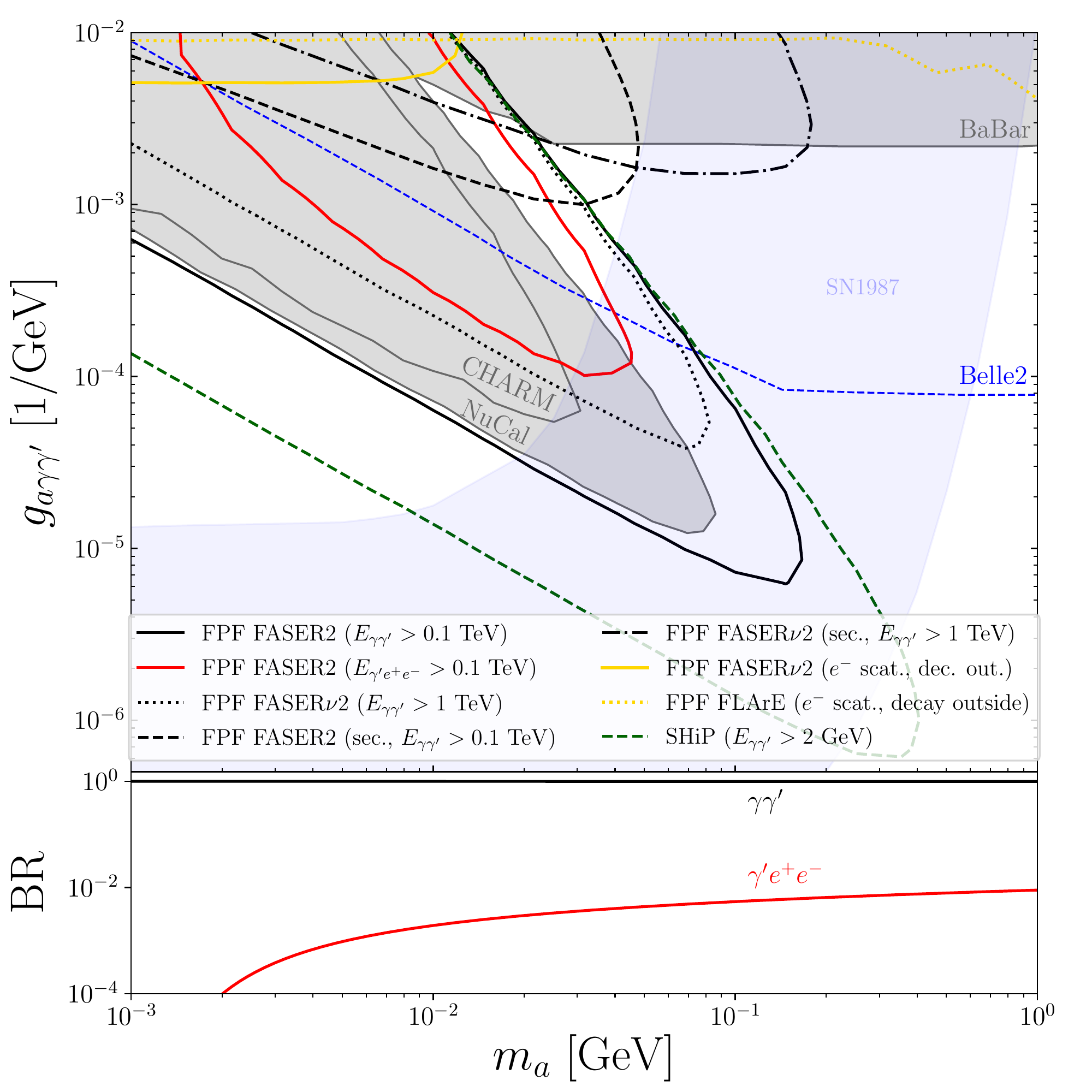}
  \caption{
    Same as \cref{fig:results_dark_photon}, but for the dark axion acting as the LLP, while the dark photon is massless. We only show the results for this mass scheme, since the results for other mass schemes are analogous to those shown in \cref{fig:results_dark_photon_mass_ratio_0.5_0.9}.
    The light-gray areas are excluded by astrophysical and cosmological bounds obtained in \cite{Hook:2021ous} for both massive and massless dark axion.
  }
  \label{fig:results_dark_axion}
\end{figure*}

\begin{figure}[tb]
  \centering
  \includegraphics[width=0.48\textwidth]{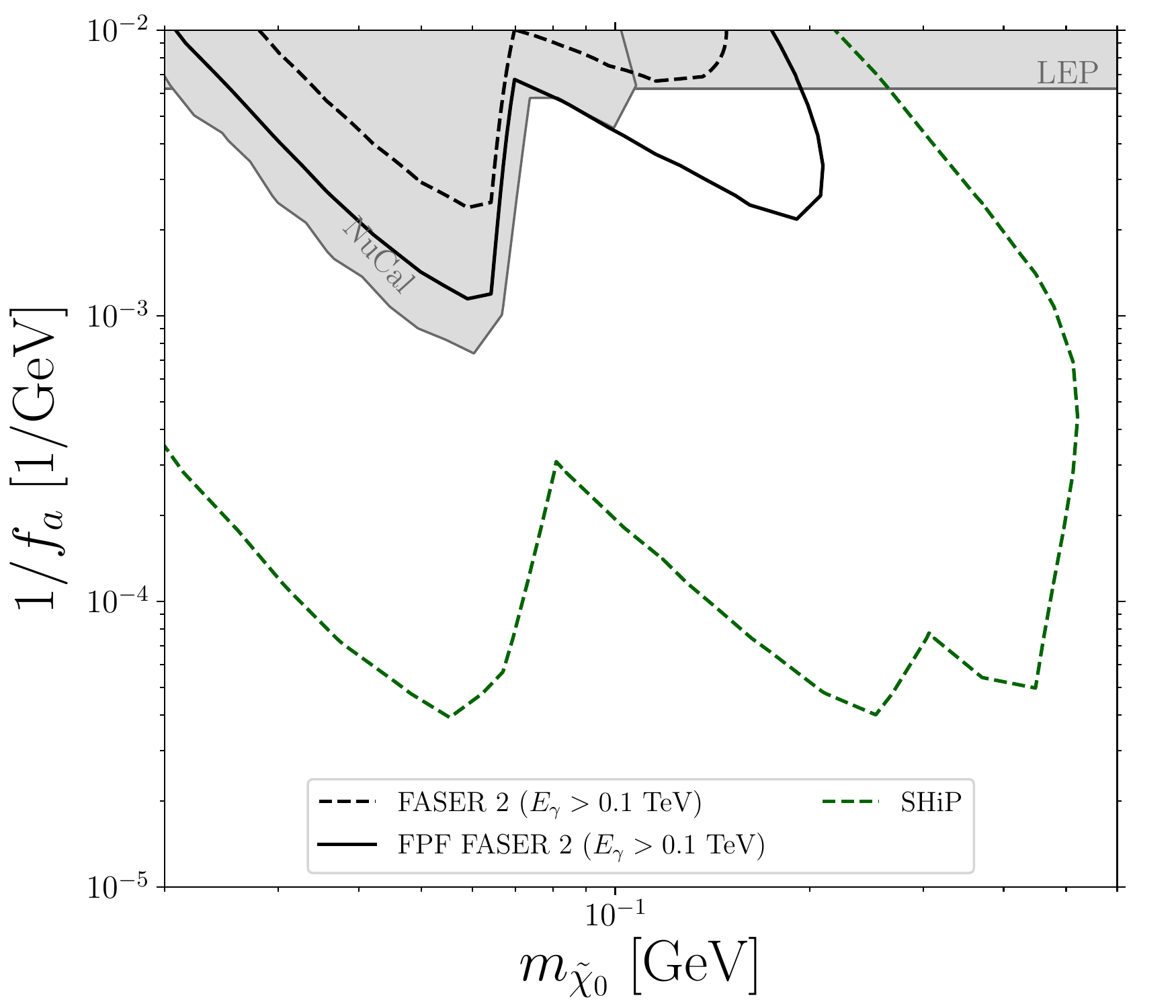}
  \caption{
    The sensitivity of FASER2 to neutralino decays into ALPino and photon for fixed $m_{\tilde{a}}=10\,\mev$. 
    The FPF version of the detector will exceed the current bounds set by NuCal and LEP due to its larger size compared to the baseline version of FASER2.
  }
  \label{fig:results_alpino}
\end{figure}

\begin{figure*}[tb]
  \centering
  \includegraphics[width=0.48\textwidth]{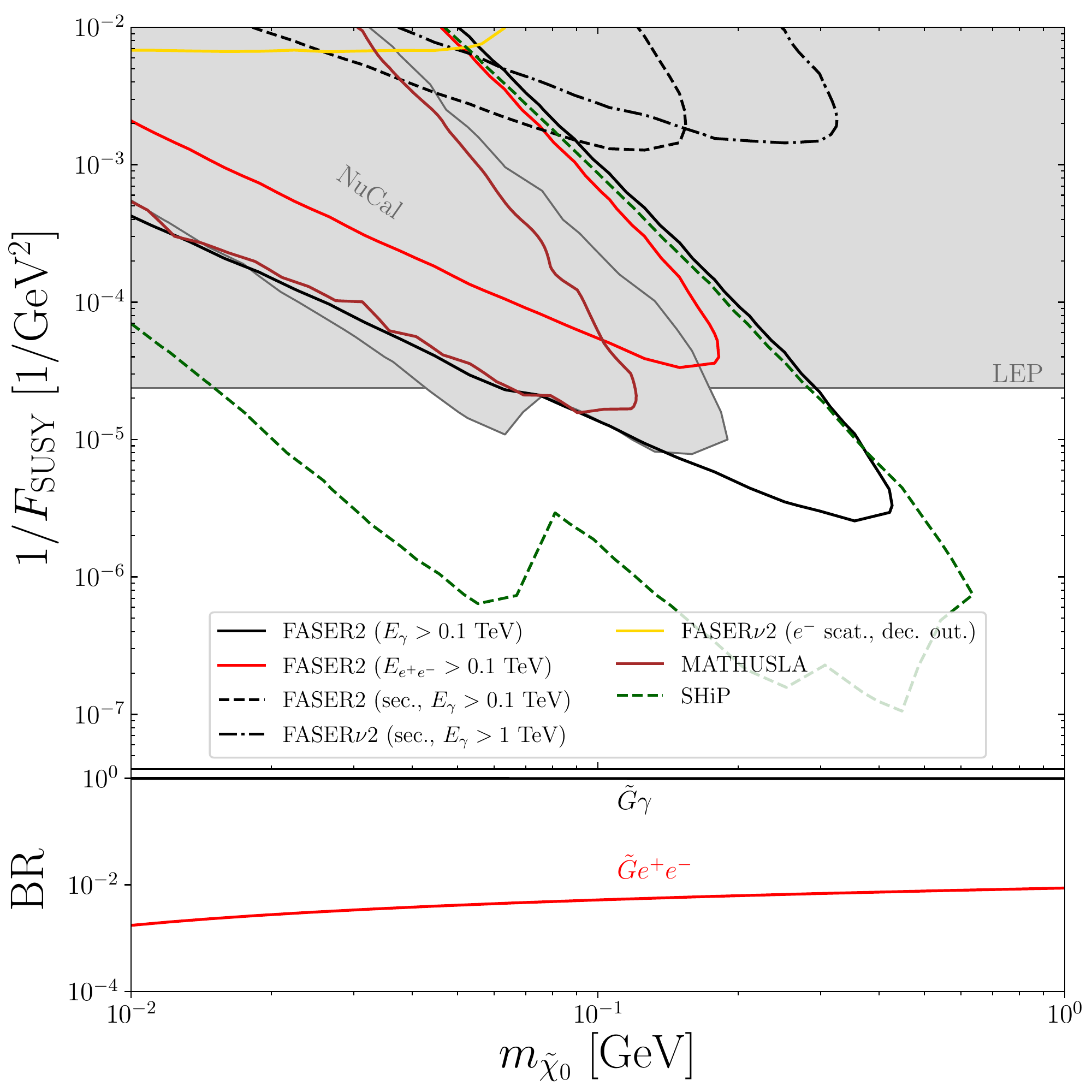}\hspace*{0.4cm}
  \includegraphics[width=0.48\textwidth]{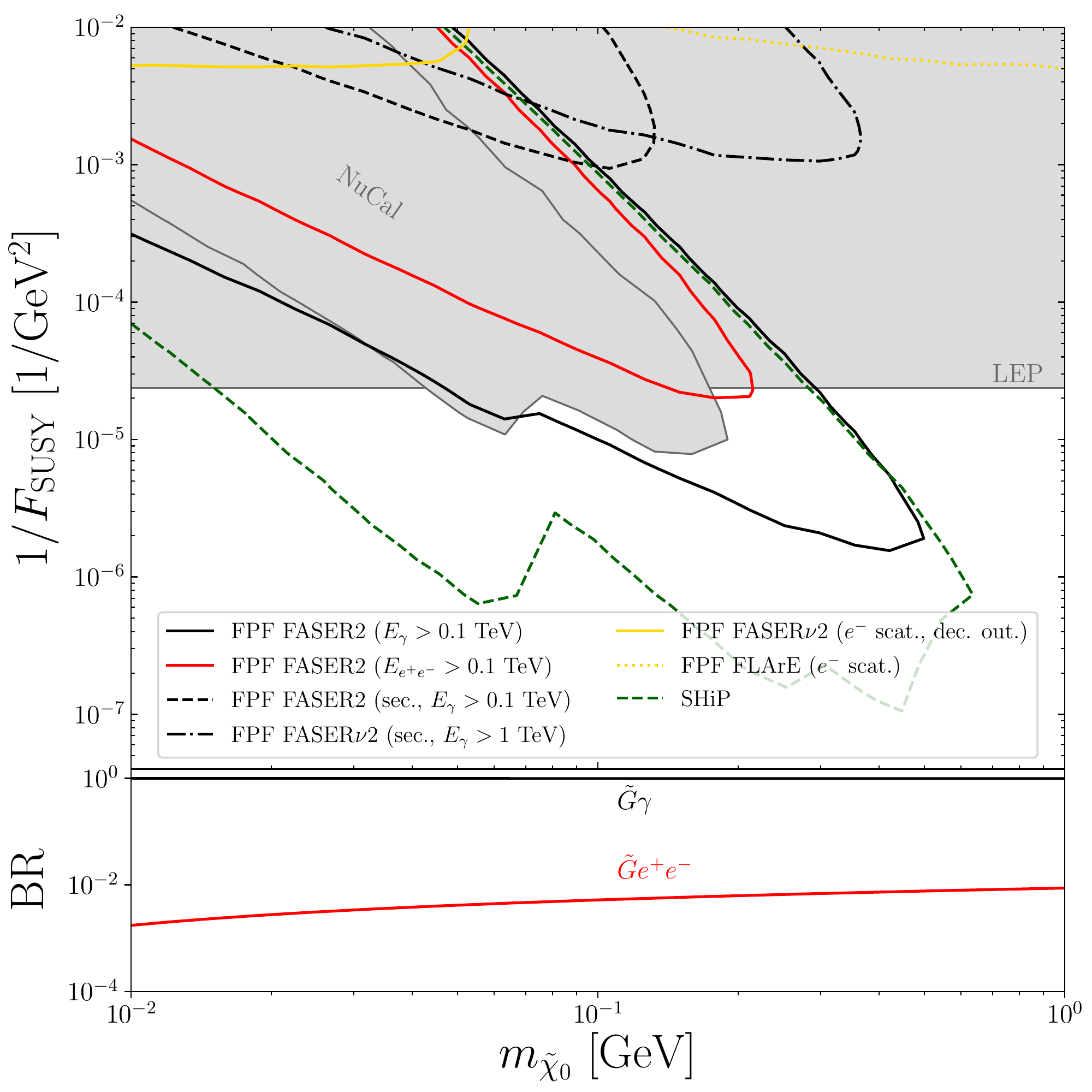}
  \caption{
    The sensitivity of FASER2, MATHUSLA, and SHiP to the neutralino-gravitino model.
    Leading two-body decays will allow FASER2 (solid black line) to partly extend the LEP bound. 
    We also present results for three-body neutralino decays at MATHUSLA (brown) and FASER2 (red solid line), which cover high and low $p_T$ regimes of LLPs produced due to the p-p collisions at the LHC, respectively.
    Secondary neutralino production extends the sensitivity of FASER2 (black dashed and dot-dashed lines) into the short-lived, higher mass regime, while electron scattering at FASER$\nu$2 and FLArE (solid and dotted gold lines, respectively) covers the lower mass regime, which, however, are both already excluded by LEP.
  }
  \label{fig:results_gravitino}
\end{figure*}

\section{Results\label{sec:results}}
In this section we present the main results of the paper: i) closed form for the Primakoff-like process, $\gamma N \to G N$, which is the main $G$ production channel, and the sensitivity reach of each experiment listed in \cref{tab:experiments} for the previously introduced LLP scenarios.

\subsection{Primakoff production of $G$ \label{sec:prod_Prim}}
The Primakoff process \cite{Primakoff:1951iae} takes place by coherent scattering with the nucleus, therefore it is restricted to the small momentum transfer regime $|t|\,\lesssim 1\, \gev^2$, where $t\equiv (p_2-p_4)^2<0$ is the $t$-channel Mandelstam variable, and $p_2$ ($p_4$) is the initial (final) momentum of a nucleus.

Following the steps described in \cref{app:sigma_Prim}, we obtained the following closed-form of this process:
\begin{dmath}[labelprefix={eq:}]
  \left(\sigma_{\gamma N \to G N}\right)_{\mathrm{univ.}} \simeq \frac{\alpha_{\mathrm{EM}} g_\gamma^2 Z^2}{2} \left(\log \left(\frac{d}{1/a^2 - t_{\mathrm{max}}}\right) - 2\right),
  \label{eq:Prim_univ}
\end{dmath}
where $g_{a\gamma\gamma}$ is the coupling between two photons and an ALP, $a=111 Z^{-1/3}/m_e$ and $d=0.164\, \gev^2 A^{-2/3}$, where $m_e$ is the electron mass, and $Z$ ($A$) is the atomic number (weight) of a nucleus. This closed-form is accurate to $\sim 1\%$ level, see \cite{Dusaev:2020gxi} for the photons ALP case.
It can be seen that for $m_G \gtrsim 5\, \gev$, $t_{\mathrm{max}}=-m_G^4/(4 E_\gamma^2)$ \cite{Feng:2018pew} becomes large, which violates the necessary coherent scattering condition, as indicated by the denominator of the logarithm. 
Technical details about the Primakoff process, including the form of the form factor given by \cref{eq:form_fact}, can be found in \cref{app:sigma_Prim}; see also discussion in \cite{Dusaev:2020gxi,Feng:2018pew}.

Then, the number of $G$ produced by in this way is
\be
N_{G} = N_\gamma \frac{\sigma_{\gamma N \to G N}}{\sigma_\gamma},
\label{eq:photon_conv_G}
\ee
where $N_\gamma$ is the number of off-shell photon and $\sigma_\gamma$ corresponds to the total cross-section of photon absorption, which for all nuclei can be found in the PDG \cite{Workman:2022ynf}.

The top left panel of \cref{fig:LLP_prod} shows the contribution of $G$ production modes, in particular the Primakoff process is indicated by the green line.
As can be seen, it is more efficient than vector meson decays, which are another leading contribution, by more than an order of magnitude.

The other panels of \cref{fig:LLP_prod} present analogous dependency of the yields as a function of LLP mass for: DAP (top right), bino-ALPino (bottom left), and bino-gravitino (bottom right). In all cases, decays of the heaviest vector meson produced in sufficiently large quantity provide the leading production mode.

\subsection{Sensitivity reaches\label{sec:results_plots}}

\paragraph{Massive spin-2 mediator}
In \cref{fig:results_univ,fig:results_nonuniversal} we present the results of our simulations for universal and photophilic coupling, respectively.

In the former case, one can notice that the coverage of the parameter space is similar to the photon-coupled ALP; see \cite{AxionLimits}.
However, the reach of each detector described in \cref{tab:experiments} is generally greater than the one for ALPs.
This is in particular evident for detectors sensitive only to decays into charged particles, such as MATHUSLA \cite{Chou:2016lxi,Curtin:2018mvb}, which can nevertheless probe the spin-2 portal with universal couplings, whereas for photophilic ALP they are only sensitive to its suppressed three-body decays.

For other detectors, which are sensitive to LLP decays into a photon pair, the increased reach is caused by a more efficient Primakoff conversion by a factor of $4$.

As mentioned earlier, the proposed extension of the FASER detector, FASER2, which is planned to take data during the High-Luminosity era of the LHC, has been proposed to be hosted in two alternative locations, with much larger size in the FPF variant \cite{MammenAbraham:2020hex,Anchordoqui:2021ghd,Feng:2022inv}; see \cref{tab:experiments}.
Moreover, in the latter scenario, the FPF will host a number of other detectors, including FLArE \cite{Batell:2021blf}.
In \cref{fig:results_univ,fig:results_nonuniversal}, we therefore consider both versions, with the results for the original FASER2 proposal on the left and the FPP results on the right.
The lines obtained by the missing energy $G$ searches at BaBar, NA64, LDMX, and M3 were taken from \cite{Voronchikhin:2022rwc}.

We note that similarly to the ALP case, SHiP will obtain the strongest bounds, followed by NA62, SeaQuest, and FASER2. Moreover, the $G\to \gamma$ search at FASER$\nu$2 and FLArE will allow to probe the low mass region.

This search is particularly important for the results for photophilic coupling shown in \cref{fig:results_nonuniversal}. Both FASER$\nu$2 and FLArE sensitivities diverge for low $m_G$ masses, due to the behavior of \cref{eq:Prim_nonuniv}.
We only plot the BaBar limit obtained in \cite{Kang:2020huh} for high mass regime, which has the same limit in both universal and photophilic coupling scenarios, leaving the study of sensitivity in the low mass regime for further work.

Moreover, the sensitivity lines obtained by $G$ decays saturate for $m_G \lesssim 50\,\mev$ regime.
It is caused by the interplay between the $\sigma \propto 1/m_G^4$ behavior of the Primakoff process producing $G$, and the $1/d \propto m_G^4$ term in \cref{eq:NoE}, which, as can be seen from \cref{eq:p_prim,eq:NoE}, cancel out.

On the other hand, the obtained sensitivity lines for photophilic coupling in the high mass regime are stronger than in the universal coupling case.
It is mainly caused by the larger $G$ lifetime and by the fact that in the universal coupling case, the branching ratio of decays to a pair of photons saturate to $\sim 0.5$, instead of being 1 as for the photophilic case.

\paragraph{Dark ALP}
When $m_{\gamma^{\prime}} > m_a$, dark photon is the LLP, and its decay width is described by \cref{eq:Gamma_two_body}.
The signatures described in \cref{sec:LLP_prod_and_sign} were simulated in modified version of $\tt FORESEE$, and the results are shown for the following mass ratios, $m_{a}/m_{\gamma^\prime}$: $0$ - \cref{fig:results_dark_photon} - and $0.5$, $0.9$ - top and bottom of \cref{fig:results_dark_photon_mass_ratio_0.5_0.9}, respectively.

In the first case, we checked that when the dark photon is produced only by the three-body pseudoscalar meson decays, we reproduce the results of \cite{deNiverville:2018hrc,deNiverville:2019xsx}.
Moreover, for the case of massless dark axion (and also in the opposite case of massless dark photon), we denote with the light-gray color the areas that are excluded by astrophysical and cosmological bounds obtained in \cite{Hook:2021ous}; see also \cite{Kalashev:2018bra,Carenza:2023qxh}.

The richness of the DAP is particularly illustrated by the top and bottom plots of \cref{fig:results_dark_photon_mass_ratio_0.5_0.9}. 
They show that when the masses of the two DS particles are comparable, the LLP decay width is suppressed and, as a result, its lifetime is longer, resulting in shifting the significant reach of FASER2 and SHiP towards higher masses.
Note that in this scenario the existing bounds, especially from NuCal, are relaxed due to the high energy threshold on the single photon, which is more difficult to meet because of the compressed spectra.
On the other hand, FASER2 reach weakens only mildly because of the typical high energy $\sim O(100)'s\,\gev$ of the produced LLPs.

Another feature of DAP that distinguishes it from the photophilic ALP is that vector meson decays produce a pair of dark photon-dark axion, both of which can travel virtually undisturbed from the production point to FASER$\nu$2, which allows for the secondary LLP production by Primakoff-like upscattering\footnote{By the same token, it also allows to study the electron scattering signature.}.
As a result, this production mode allows to cover part of the $d=\gamma c\tau \sim 1\,\m$ region of the parameter space, see dashed and dash-dotted lines in \cref{fig:results_dark_photon,fig:results_dark_photon_mass_ratio_0.5_0.9,fig:results_dark_axion}.
Note that the probability of LLP decay taking place inside the decay vessel in short-lived regime is exponentially suppressed, $p(E) \simeq e^{-L/d}$ for $d \ll L$, hence this region of the parameter space cannot be covered by a detector placed at a significant distance from the LLP production point.

Lastly, the electron scattering signature allows coverage of the low-mass regime and is complementary to the decays of the dark photons produced in both primary and secondary production processes.
It should be noted that the electron scattering limit is typically weaker than in the case of secondary production, mainly due to the lack of $Z^2$ enhancement, cf. \cref{eq:Prim} and \cref{eq:e_scat}.

The results for the opposite mass hierarchy are shown in \cref{fig:results_dark_axion}.
The formulas for the LLP production channels are the same as for the case of the dark photon acting as the LLP, while the LLP lifetime is smaller by a factor of $3$; see \cref{eq:Gamma_two_body}.
As a result, the sensitivity lines are shifted towards smaller masses.
Moreover, the Primakoff and electron scattering cross-sections are also smaller by a factor of $3$, resulting in smaller reach.

We only show one benchmark corresponding to massless dark photon, while other mass scenarios are analogous to the top and bottom rows of \cref{fig:results_dark_photon_mass_ratio_0.5_0.9}.

\paragraph{Bino}
In \cref{fig:results_alpino}, we present our results for the scenario when ALPino is the LSP.
For beam dump experiment, we find agreement with results of \cite{Choi:2019pos}. We consider additional detector of this type, NuCal, and we find that it actually improves over the NOMAD \cite{NOMAD:1997pcg} sensitivity shown in that work.

Since the leading channel of NLSP-LSP production is $f_a$-independent meson decay into a pair of binos, there is hardly any flux of ALPinos - see the bottom left panel of \cref{fig:LLP_prod}.
As a result, neither the secondary production given by the top line of \cref{eq:Prim_bino}, nor upscattering on electrons, given by the top line of \cref{eq:Prim_e}, are efficient.
Consequently, FASER2 will not have sensitivity to such signatures.

On the other hand, bino-pair production can be quite efficient.
While the baseline versions of FASER2 taking data during the High Luminosity era of the LHC will not improve over NuCal (but its sensitivity is greater than NOMAD), the FPF FASER2 will extend it in the $m_{\tilde{\chi}_0} \simeq 0.1\,\gev$ mass regime.
For LLP decays produced in the primary production, its main advantage over the baseline version is simply its larger size.
Finally, we checked that the three-body decays do not lead to sensitivity in the allowed region of parameter space for any of the detectors considered.

On the other hand, when gravitino acts as the LSP, the dominant production modes produce equal fluxes of gravitinos and neutralinos, allowing the additional upscattering signatures described in \cref{sec:LLP_prod_and_sign}.
In fact, contrary to the ALPino scenario, both neutralino production and decay processes are controlled by the NLSP-LSP-photon coupling, which here depends on the SUSY breaking scale as $1/\sqrt{F_{\mathrm{SUSY}}}$.

This allows one to search not only for the displaced $\tilde{\chi}_0$ decays, but also for the electron scattering signature and for the decays of $\tilde{\chi}_0$ produced by upscattering occuring at the FASER$\nu$2 detector located before FASER2.
In \cref{fig:results_gravitino}, we present our main results for this model.
The areas shaded in gray are excluded by NuCal or LEP \cite{DELPHI:2003dlq,Mawatari:2014cja}.

As mentioned earlier, we consider two versions of the FASER2 detector - the results for the baseline version are in the left panel, while the results for FPF FASER2 are in the right panel.
The sensitivity lines derived for the two-body bino decays are marked by black lines for FASER and green for SHiP, while those for three-body decays are indicated by red (FASER2) and brown (MATHUSLA) lines. 
The sensitivity lines correspond to the number of bino decays (number of LLP signatures in the general case) given in Tab. 1 in \cite{Jodlowski:2023sbi} for each detector considered.

As is clearly seen, FASER2 will be able to significantly extend the LEP limit for $m_{\tilde{\chi}_0} \gtrsim 0.1\,\gev$ mass range, while searches for $e^+ e^-$ pairs produced in the three-body decays at MATHUSLA and FASER2 will be competitive with current LEP and NuCal bounds.
Moreover, the FPF version of FASER2 and SHiP may improve them even further. 

The upscattering signatures allow to cover the smaller lifetime regime, $d_{\tilde{\chi}_0}\sim 1\,\m$, which, however, is already excluded by LEP for both locations of the bino decays: FASER2 (black dashed line) and FASER$\nu$2 (black dot-dashed line).
Finally, the electron scattering signature at FASER$\nu$2 and FLArE (gold solid and dot-dashed line, respectively) covers the low mass region of the bino, which, is also already excluded.

\section{Conclusions\label{sec:conclusions}}
Photon-coupled LLPs are well motivated extensions of the SM and constitute a prime target for beam dump and other intensity frontier experiments, in particular FASER2, FASER$\nu$2, MATHUSLA, NA62, SeaQuest, and SHiP.
In this paper we have investigated the prospects of testing several such models, among them a massive spin-2 particle with (non-)universal couplings to the SM fields.
We have shown that although the considered model bears similarities to the well-established photophilic ALP, several important differences occur. These include i) a more efficient Primakoff conversion by a factor of $4$, ii) a smaller decay width by a factor of $0.8$, and iii) the loss of perturbative unitarity for the non-universal couplings to the SM fields in the $m_G \lesssim 50\,\mev$ regime for the Primakoff process, $\sigma \propto 1/m_G^4$, due to non-decoupling of the helicity $0$ states of the massive spin-2 particle.
We found that SHiP will provide the strongest bounds, reaching up to $m_G=1.7\,\gev$ for $g_{\gamma\gamma} \gtrsim 4.8 \times 10^{-8} \,\gev^{-1}$ ($m_G=2.1\,\gev$ for $g_{\gamma\gamma}\gtrsim 3.6 \times 10^{-8} \,\gev^{-1}$) for universal (non-universal) $G$ coupling reaching significantly below the current bounds obtained from BaBar and NuCal.

We have also studied LLPs coupled to a single photon, among them the dark axion portal and a light neutralino coupled to ALPino or gravitino.
The main difference between single- and two-photon couplings is that the Primakoff conversion of an on-shell photon into a LLP is no longer possible, and the leading LLP production modes  are vector meson decays, yielding approximately an order of magnitude fewer LLPs.
Another challenge is that the LLP decays semi-visibly, so its energy can be deposited almost exclusively by a single high-energy photon. 
Such an experimental signature is more challenging than the usual two-photon or two-lepton LLP decay because of an additional SM induced background.
On the other hand, future detectors like FASER2 and SHiP will be able to effectively probe such LLP decays, resulting in sizable coverage of the parameter space for the DAP and sub-GeV bino coupled ALPino or gravitino.

Moreover, secondary LLP production taking place just in front of the decay vessel will allow covering part of the shorter LLP lifetime regime corresponding to $d\sim 1\,\m$. 
This is in contrast to photophilic ALP, for which such process is impossible because a photon would be promptly absorbed after its production.
Moreover, the electron scattering signature for ALP is also challenging, as it would lead to both electron recoil and a single high-energy photon, which would typically be vetoed. 
Therefore, the extended DS content of BSM scenarios predicting single-photon coupled LLPs allows one to probe them by such more extensive signatures.
What is more, the low mass LLP regime, $m\lesssim 10\,\mev$, can be studied by scatterings of either of the DS species with electrons taking place inside FASER$\nu$2 or FLArE.
This is a complementary search to both the LLP displaced decays and missing energy searches, which are limited to nearly disjoint LLP mass ranges.
Finally, we considered the extended version of the FASER2 experiment, the proposed FPF. 
Due to its larger size and expanded capabilities, such a facility could significantly expand the limits for all LLP scenarios considered, being competitive with SHiP, which, due to its higher luminosity, might further improve the limits.

\acknowledgments
This work was supported by the Institute for Basic Science under the project code, IBS-R018-D1.

\appendix
\section{LLP decays}
\label{app:decays}

Below, we give the relevant decay widths. For each case when only the leading form of the expression is given, its full form, which we used in our simulations, can be found in the Mathematica notebook included in the auxiliary materials.

\paragraph{Massive spin-2 mediator}
The widths of $G$ decays into a pair of photons or SM leptons are \cite{Giudice:1998ck,Han:1998sg,Lee:2013bua}
\begin{dmath}[labelprefix={eq:}]
  {\Gamma_{G \to \gamma\gamma} = \frac{g_{\gamma\gamma}^2 m_G^3}{80\pi},} \\
  {\Gamma_{G \to l^+ l^-} = \frac{g_l^2 m_G^3}{160\pi} \left(1 - \frac{4 m_l^2}{m_G^2}\right)^{3/2} \left(1 + \frac{8 m_l^2}{3 m_G^2}\right).}
  \label{eq:Gamma_G2}
\end{dmath}

\paragraph{Dark ALP}
The decay widths for the two-body final states are \cite{Kaneta:2016wvf}
\begin{dmath}[labelprefix={eq:}]
  {\Gamma_{\gamma^{\prime} \to \gamma a} = \frac{g_{a\gamma\gamma^{\prime}}^2}{96 \pi} m_{\gamma^{\prime}}^3 \left(1-\frac{m^2_a}{m^2_{\gamma^{\prime}}}\right)^3,} \\
  {\Gamma_{a\to \gamma \gamma^{\prime}} = \frac{g_{a\gamma\gamma^{\prime}}^2}{32 \pi} m_a^3 \left(1-\frac{m^2_{\gamma^{\prime}}}{m^2_a}\right)^3.}
  \label{eq:Gamma_two_body}
\end{dmath}

Since some LLP detectors may not be sensitive to a single-photon decays, we also considered phase-space suppressed three-body decays of a dark photon and a dark ALP, which are described be the following expressions in the $m_{\gamma^{\prime}} \gg m_a, m_l$ and $m_a \gg m_{\gamma^{\prime}}, m_l$ limits, respectively:
\begin{widetext}
  \begin{dmath}[labelprefix={eq:}]
    \Gamma_{\gamma^{\prime}\to l^+ l^- a} = \frac{\alpha_{\mathrm{EM}} g_{a\gamma \gamma^{\prime}}^2}{576 \pi ^2 m_{\gamma}^3} \left(32 m_l^6 \coth^{-1}\left(\frac{m_{\gamma^{\prime}}}{\sqrt{m_{\gamma^{\prime}}^2-4 m_l^2}}\right) + m_{\gamma^{\prime}} \left(\sqrt{m_{\gamma^{\prime}}^2-4 m_l^2} \left(26 m_{\gamma^{\prime}}^2 m_l^2-7 m_{\gamma^{\prime}}^4+8 m_l^4\right)-4 m_{\gamma^{\prime}}^5 \log \left(\frac{2 m_l}{\sqrt{m_{\gamma^{\prime}}^2-4 m_l^2}+m_{\gamma^{\prime}}}\right)+12 m_{\gamma^{\prime}} m_l^4 \log \left(\frac{16 m_l^4 \left(m_{\gamma^{\prime}}-\sqrt{m_{\gamma^{\prime}}^2-4 m_l^2}\right)}{\left(\sqrt{m_{\gamma^{\prime}}^2-4 m_l^2}+m_{\gamma^{\prime}}\right)^5}\right)\right)\right),
    \label{eq:gprime_lla}
  \end{dmath}
\end{widetext}

\begin{widetext}
  \begin{dmath}[labelprefix={eq:}]
    \Gamma_{a\to l^+ l^- \gamma^{\prime}} = \frac{\alpha_{\mathrm{EM}} g_{a\gamma \gamma^{\prime}}^2}{192 \pi ^2 m_a^3} \left(32 m_l^6 \coth^{-1}\left(\frac{m_a}{\sqrt{m_a^2-4 m_l^2}}\right) + m_a \left(\sqrt{m_a^2-4 m_l^2} \left(26 m_a^2 m_l^2-7 m_a^4+8 m_l^4\right)-4 m_a^5 \log \left(\frac{2 m_l}{\sqrt{m_a^2-4 m_l^2}+m_a}\right)+12 m_a m_l^4 \log \left(\frac{16 m_l^4 \left(m_a-\sqrt{m_a^2-4 m_l^2}\right)}{\left(\sqrt{m_a^2-4 m_l^2}+m_a\right)^5}\right)\right)\right).
    \label{eq:a_llg}
  \end{dmath}
\end{widetext}

\paragraph{Bino}

The two-body decay width for bino decaying into an ALPino and a photon is \cite{Choi:2019pos}
\be
  \Gamma_{\tilde{\chi}_0 \to \tilde{a}\gamma} = \frac{\alpha_{\mathrm{EM}}^2 \cos^2\theta_W}{128 \pi^3} \frac{m_{\tilde{\chi}_0}^3}{f_a^2} \left(1-\frac{m_{\tilde{a}}^2}{m_{\tilde{\chi}_0}^2}\right)^3,
  \label{eq:Gamma_axino_2body}
\ee
while the decay width for the leading three-body decay into an ALPino and an electron-positron pair in the limit of $m_{\tilde{\chi}_0} \gg m_{\tilde{G}}, m_{e^-}$ is
\begin{dmath}[labelprefix={eq:}]
  \Gamma_{\tilde{\chi}_0 \to \tilde{a} e^+ e^-} \simeq \frac{\alpha_{\mathrm{EM}}^3 \cos^2\theta_W}{1152 \pi^4 f_a^2 m_{\tilde{\chi}_0}^3}\\ \times {\left(18 m_{\tilde{\chi}_0}^4 m_{e^-}^2 - 4 m_{\tilde{\chi}_0}^6 - 32 m_{e^-}^6 + 3 m_{\tilde{\chi}_0}^6 \log \left(\frac{m^2_{\tilde{\chi}_0}}{4 m_{e^-}^2}\right) \right).}
  \label{eq:Gamma_axino_3body}
\end{dmath}

The two-body decay width for bino decaying into a gravitino and a photon is
\be
  \Gamma_{\tilde{\chi}_0 \to \tilde{G}\gamma} =& \frac{\cos^2\theta_W m^5_{\tilde{\chi}_0}}{16 \pi F_{\mathrm{SUSY}}^2} \left(1-\frac{m^2_{\tilde{G}}}{m^2_{\tilde{\chi}_0}}\right)^3 \left(1+\frac{m^2_{\tilde{G}}}{m^2_{\tilde{\chi}_0}}\right).
  \label{eq:Gamma_grav_2body}
\ee
We used the Feynman rules described in \cite{Pradler:2006tpx}. In particular, we used the full form of the gravitino polarization tensor, which is defined as the sum of the gravitino field with momentum $p$ over its spin degrees of freedom, 
\be
  \Pi^{\pm}_{\mu\nu}(k) \equiv \sum_{s=\pm\frac 12, \pm\frac 32} \psi^{\pm, s}_\mu(k) \overline{\psi}^{\pm, s}_\nu(k).
  \label{eq:grav_tensor}
\ee
In the high-energy limit, where ${\Pi^{\pm}_{\mu\nu}(k) \simeq -\slashed{k}\,(g_{\mu\nu} - 2p_\mu p_\nu/3m^2_{\tilde{G}}})$, we match the well-known result \cite{Ellis:2003dn,Giudice:1998bp,Diaz-Cruz:2016abv}, 
\be
  \Gamma_{\tilde{\chi}_0 \to \tilde{G}\gamma} = \frac{\cos^2\theta_W m^5_{\tilde{\chi}_0}}{16 \pi F_{\mathrm{SUSY}}^2} \left(1-\frac{m^2_{\tilde{G}}}{m^2_{\tilde{\chi}_0}}\right)^3 \left(1+3\frac{m^2_{\tilde{G}}}{m^2_{\tilde{\chi}_0}}\right).
\ee

In the limit of $m_{\tilde{\chi}_0} \gg m_{\tilde{G}}, m_{e^-}$, the bino decay into a gravitino and an electron-positron pair is described by the following formula:
\be
  \Gamma_{\tilde{\chi}_0 \to \tilde{G} e^+ e^-} \simeq & \frac{\alpha_{\mathrm{EM}} \cos^2\theta_W m_{\tilde{\chi}_0}^5}{576 \pi^2 F_{\mathrm{SUSY}}^2} \times \\
  & \left(24 \log \left(\frac{m_{\tilde{\chi}_0}}{m_{e^-}}\right) - 25 - 12 \log (4)\right).
  \label{eq:Gamma_grav_3body}
\ee

\section{Vector meson decays}
\label{app:prod_vec}

The following are formulas for vector meson decays mediated by an off-shell photon that result in the production of LSP-NLSP pair, $V(p_0) \!\to\! \gamma^*(p_1+p_2) \!\to\! \mathrm{LSP}(p_1) + \mathrm{NLSP}(p_2)$.

\paragraph{Massive spin-2 mediator}
\begin{dmath}[labelprefix={eq:}]
      \left(\frac{{\rm BR}_{V \rightarrow \gamma G}}{{\rm BR}_{V \rightarrow e^+ e^-}}\right)_{\mathrm{univ.}} \!=\! \frac{ g_{\gamma\gamma}^2 \left(M^2-m_G^2\right)^3}{8 \pi  \alpha_{\text{EM}} M \sqrt{M^2-4 m_e^2} \left(M^2+2 m_e^2\right)},
\end{dmath}
\begin{dmath}[labelprefix={eq:}]
  \left(\frac{{\rm BR}_{V \rightarrow \gamma G}}{{\rm BR}_{V \rightarrow e^+ e^-}}\right)_{\mathrm{non. univ.}} \!=\! \frac{ g_{\gamma\gamma}^2 \left(M^2-m_G^2\right)^3 (3 m_G^2 M^2 + 6 m_G^4 + M^4)}{16 \pi  \alpha_{\text{EM}}  M m_G^4 \sqrt{M^2-4 m_e^2} \left(M^2+2 m_e^2\right)},
\end{dmath}
where ${\rm BR}_{V\rightarrow e^+ e^-}$ is the branching ratio corresponding to decays into $e^+ e^-$ \cite{Workman:2022ynf}, which we took from the PDG \cite{Workman:2022ynf}.

We also checked that for the non-universal coupling of $G$ to a pair of photons, the three-body decays of pseudoscalar mesons, $P(p_0) \!\to\! \gamma(p_1)+ \gamma^*(p_2+p_3) \!\to\! \gamma(p_1) + \gamma(p_2) + G(p_3)$, provide subleading contribution relative to the decays of vector mesons. We expect the same to hold for the universal coupling case.

\paragraph{Dark ALP}
\be
  \frac{{\rm BR}_{V \rightarrow a \gamma^\prime}}{{\rm BR}_{V \rightarrow ee}} = \frac{ g_{a\gamma\gamma^\prime}^2 \left((-M^2+m_a^2+m_{\gamma^\prime}^2)^2-4 m_a^2 m_{\gamma^\prime}^2\right)^{3/2}}{32 \pi  \alpha_{\text{EM}} M \sqrt{M^2-4 m_e^2} \left(M^2+2 m_e^2\right)}.
  \label{eq:br_vec}
\ee

\paragraph{Bino}
\phantom{aa}
\begin{widetext}
  \be
    \label{eq:brV}
    \frac{{\rm BR}_{V \rightarrow \tilde{a}\tilde{\chi}_0}}{{\rm BR}_{V \rightarrow e^+ e^-}} \!=\! \cos^2\theta_W & \frac{\alpha_{\text{EM}} \left(m_V^2+2 (m_{\tilde{a}}-m_{\tilde{\chi}_0})^2\right) (m_V^2-(m_{\tilde{a}}+m_{\tilde{\chi}_0})^2) \sqrt{\left(-m_V^2+m_{\tilde{a}}^2+m_{\tilde{\chi}_0}^2\right)^2-4 m_{\tilde{a}}^2 m_{\tilde{\chi}_0}^2}}{128 \pi ^3 f_a^2 \sqrt{m_V^2-4 m_e^2} \left(m_V^3+2 m_V m_e^2\right)}, \\
    \frac{{\rm BR}_{V \rightarrow \tilde{G} \tilde{\chi}_0 }}{{\rm BR}_{V \rightarrow e^+ e^-}} \!=\! \cos^2\theta_W & \frac{ (m_V^2 - (m_{\tilde{G}}+m_{\tilde{\chi}_0})^2) \sqrt{\left(-m_V^2+m_{\tilde{G}}^2+m_{\tilde{\chi}_0}^2\right)^2-4 m_{\tilde{G}}^2 m_{\tilde{\chi}_0}^2}}{8 \pi F_{\mathrm{SUSY}}^2 \alpha_{\text{EM}} \sqrt{m_V^2-4 m_e^2} \left(m_V^3+2M  m_e^2\right)} \times \\
    & \times \left(2 m_V^2 \left(m_{\tilde{G}}^2+m_{\tilde{G}} m_{\tilde{\chi}_0}-m_{\tilde{\chi}_0}^2\right)+m_V^4+(m_{\tilde{G}}-m_{\tilde{\chi}_0})^2 \left(3 m_{\tilde{G}}^2+m_{\tilde{\chi}_0}^2\right)\right).
  \ee
\end{widetext}

\section{Pseudoscalar meson decays}
\label{app:d2Br}
Another LLP production mode, which is typically subdominant to Primakoff process and vector meson decays (when these are applicable), are decays of pseudoscalar mesons into a photon and DS states mediated by an off-shell photon, $P(p_0) \!\to\!  \gamma(p_1)+ \gamma^*(p_2+p_3) \!\to\!  \gamma(p_1) + a(p_2) + \gamma^{\prime}(p_3)$.

\paragraph{Dark ALP}
We obtained the same averaged amplitude squared as \cite{deNiverville:2018hrc}, while below we give the resulting differential branching ratio in a form convenient for Monte Carlo simulation:
\begin{widetext}
\be
  \frac{d{\rm BR}_{P \rightarrow \gamma a \gamma^{\prime}}}{dq^2 d\cos\theta} = {\rm BR}_{P\rightarrow \gamma \gamma}  &\!\times \!\! \left[ \frac{g_{a\gamma\gamma^\prime}^2}{256 \pi ^2 m_P^6 q^6} \left(m_P^2-q^2\right)^3 (\cos(2\theta)+3) \left( (m_{\gamma^\prime}^2 + m_a^2 - q^2)^2 - 4 m_{\gamma^\prime}^2  m_a^2 \right)^{3/2}\!\right],
  \label{eq:br2dq2dcostheta}
\ee
\end{widetext}
where $m_P$ is the pseudoscalar meson mass, $q^2 \equiv (p_2+p_3)^2$ is the momentum squared of the off-shell photon mediating the decay, and $\theta$ is the angle between between the LLP momentum in the rest frame of the off-shell photon and the momentum of the off-shell photon in the meson rest frame; ${\rm BR}_{P\rightarrow \gamma \gamma}$ is the branching ratio of pseudoscalar meson decaying into two photons taken from the PDG \cite{Workman:2022ynf}.

\paragraph{Bino}
\phantom{aaa}
\begin{widetext}
  \be
    \label{eq:br2dq2dcostheta}
    \frac{d{\rm BR}_{P \!\to\! \gamma \tilde{a} \tilde{\chi}_0}}{dq^2 d\cos\theta} = {\rm BR}_{P\rightarrow \gamma \gamma} \cos^2\theta_W &\!\times\!\! \Bigg[ \frac{\alpha_{\mathrm{EM}}^2 }{512 \pi ^4 f_a^2 m_P^6 q^6} \left(q^2 - m_P^2\right)^3 \sqrt{\left(m_{\tilde{\chi}_0}^2+m_{\tilde{a}}^2-q^2\right)^2-4 m_{\tilde{\chi}_0}^2 m_{\tilde{a}}^2} \\ 
    &\times \left((m_{\tilde{\chi}_0}+m_{\tilde{a}})^2-q^2\right) \left(\cos (2\theta) \left((m_{\tilde{\chi}_0}-m_{\tilde{a}})^2-q^2\right)+3 (m_{\tilde{\chi}_0}-m_{\tilde{a}})^2+q^2\right) \!\Bigg], \\    
    \frac{d{\rm BR}_{P \!\to\! \gamma \tilde{G} \tilde{\chi}_0}}{dq^2 d\cos\theta} = {\rm BR}_{P\rightarrow \gamma \gamma} \cos^2\theta_W &\!\times\!\! \left[ \frac{1}{64 \pi^2 F_{\mathrm{SUSY}}^2 m_P^6 q^6} (m_P^2-q^2)^3 (m_{\tilde{\chi}_0}^2-q^2)^4 (\cos (2\theta)+3) \!\right].
  \ee
\end{widetext}

\section{Primakoff upscattering cross-sections}
\label{app:sigma_Prim}

The general formula for the Primakoff process involving a particle with initial momentum $p_1$ on nucleus $N$, resulting in an outgoing particle with momentum $p_3$ and an unperturbed nucleus $N$, is given by:
\be
  \sigma = \int_{t_{\mathrm{min}}}^{t_{\mathrm{max}}} \frac{\overline{|M|}^2\, F(t)^2\, dt}{16\pi \left[(s-m_1^2-m_2^2)^2-4 m_1^2 m_2^2\right]},
  \label{eq:sigma_integ}
\ee
where $\overline{|M|}^2$ is the average of the squared amplitude of the considered process, $s \equiv (p_1+p_2)^2=m_1^2+m_2^2+2m_2 E_1$, $t=(p_1-p_3)^2<0$, and $-1\, \gev^2 \simeq t_{\mathrm{min}}<t_{\mathrm{max}} \simeq -(m_1^4+m_3^4)/(4E_1^2)$, where the last formula was derived in an analogous way as the $m_3=0$ case discussed in \cite{Feng:2018pew}.

The form factor $F(t)$ guarantees the screening of the nucleus by the atomic electrons. We consider the following momentum-dependent elastic atomic form-factor \cite{Schiff:1953yzz,Tsai:1973py,Kim:1973he}:
\be
F(-t) \equiv Z\left(\frac{a^2 t}{1+a^2 t}\right)\left(\frac{1}{1 + t/d}\right),
\label{eq:form_fact}
\ee
where $a=111 Z^{-1/3}/m_e$ and $d=0.164\, \gev^2 A^{-2/3}$, where $m_e$ is the electron mass, and $Z$ ($A$) is the atomic number (weight) of a nucleus.
The atomic form-factor effectively restricts the scattering to the small momentum transfer regime $|t|\,\lesssim 1\, \gev^2$.
We note that one could consider other forms of the form-factor, \eg, the nuclear Helm's form-factor \cite{Helm:1956zz}. However, we found that such a choice has typically only a $\mathcal{O}(1)\%$ influence on the cross-section, which is in agreement with results of \cite{Voronchikhin:2022rwc}.

Moreover, the form factor given by \cref{eq:form_fact} allows to integrate \cref{eq:sigma_integ} analytically without any approximations, which speeds up the numerical simulation.
However, the integrated expressions have a very long form, therefore we only give the leading contributions, see \cref{eq:Prim_univ,eq:Prim_nonuniv}. 

The method described in \cite{Dusaev:2020gxi} is based on power series decomposition of the differential cross-section, 
\begin{dmath}[labelprefix={eq:}]
  \frac{d\sigma_{\text{Prim.}}}{dt} = \left(\frac{a_0 + a_1 t + \dots}{t^2}\right) F(-t)^2,
  \label{eq:dsigmadt}
\end{dmath}
where $a_0$, $a_1$, $\dots$ are $t$-independent quantities, and $a_2$, $\dots$ were shown to give negligible contributions.
Then, \cref{eq:dsigmadt} was integrated over an extended interval $[-\infty,t_{\mathrm{max}}]$.

\paragraph{Massive spin-2 mediator}
For computation of $\gamma N \to G N$ cross-section for the universal, and photophilic $G$ coupling, we use results of \cite{Gill:2023kyz}, in particular Eq. 23-26 therein. 
That work considered the $l \bar{l} \to G \gamma$ process, therefore we use crossing symmetry relations to obtain the $l \gamma \to l G$ amplitude. 
Following the same steps as \cite{Dusaev:2020gxi}, we obtained \cref{eq:Prim_univ} and the following formula for the universal and photophilic coupling, respectively:
\begin{dmath}[labelprefix={eq:}]
  \left(\sigma_{\gamma N \to G N}\right)_{\mathrm{non-univ.}} \simeq  {\frac{\alpha_{\mathrm{EM}} g_\gamma^2 Z^2}{2} \bigg[ \log \left(\frac{d}{1/a^2 - t_{\mathrm{max}}}\right) - 2}  - \frac{d^2}{6 m_G^4} \left(\log(d) + 1 \right) \bigg].
  \label{eq:Prim_nonuniv}
\end{dmath}

Let us note that the large $m_G$ limit is the same for both cases, while only the latter scenario leads to the $\sigma \propto 1/m_G^4$ enhancement, which leads to the aforementioned unitarity violation \cite{Artoisenet:2013puc}.
We will also use the cross-section for the inverse process, the conversion of $G$ into a single high-energy photon, which is given by $\sigma_{G N \to \gamma N} = 2/5 \, \sigma_{\gamma N \to G N}$, coming from the photon and $G$ degrees of freedom - 2 and 5, respectively.


\paragraph{Dark ALP}
\begin{dmath}[labelprefix={eq:}]
  \sigma_{\gamma^{\prime} N \to a N} \simeq \frac{\alpha_{\mathrm{EM}} g_{a\gamma\gamma^{\prime}}^2 Z^2}{12} \left(\log \left(\frac{d}{1/a^2 - t_{\mathrm{max}}}\right)-2\right).
  \label{eq:Prim}
\end{dmath}

\paragraph{Bino}
\phantom{aa}
\be
  \sigma_{\tilde{a} N - \tilde{\chi}_0 N} \simeq & \frac{\alpha_{\mathrm{EM}}^3 \cos^2\theta_W Z^2}{16 \pi ^2 f_a^2} \times \\
  &\left(\log \left(\frac{d}{1/a^2 - t_{max}}\right) - 2\right), \\
  \sigma_{\tilde{G} N - \tilde{\chi}_0 N} \simeq & \frac{\alpha_{\mathrm{EM}} \cos^2\theta_W Z^2}{2 F_{\mathrm{SUSY}}^2} \times \\
  &\left( d + m_{\tilde{\chi}_0}^2\left(\log \left(\frac{d}{1/a^2 - t_{max}}\right) - 2\right) \right).
  \label{eq:Prim_bino}
\ee

\subsection{Electron scattering}
Below we give the formulas for the integrated cross-sections for scattering with electrons. 
The expressions for the differential cross-section, $d\sigma/dE_R$, where $E_R$ is the electron recoil energy, can be found in the Mathematica notebook. These expressions are needed to impose the angular and energy cuts indicated in \cref{tab:experiments}.

\paragraph{Dark ALP}
\begin{dmath}[labelprefix={eq:}]
  \sigma_{\gamma^{\prime} e^- \to a e^-} \simeq \frac{\alpha_{\mathrm{EM}} g_{a\gamma\gamma^{\prime}}^2}{12} \log \left( \frac{E_R^{\mathrm{max}}}{E_R^{\mathrm{min}}} \right).
  \label{eq:e_scat}
\end{dmath}

\paragraph{Bino}
\phantom{aa}
\be
  \sigma_{\tilde{a} e^- \to \tilde{\chi}_0 e^-} \simeq & \frac{\alpha_{\mathrm{EM}}^3 \cos^2\theta_W}{16 \pi^2 f_a^2} \times \log\left(\frac{E_R^{\mathrm{max}}}{E_R^{\mathrm{min}}}\right), \\
  \sigma_{\tilde{G} e^- \to \tilde{\chi}_0 e^-} \simeq & \frac{\alpha_{\mathrm{EM}}\cos^2\theta_W}{2 F_{\mathrm{SUSY}}^2} \times \\
  & \left(2 m_e (E_R^{\mathrm{max}}-E_R^{\mathrm{min}}) +  m_{\tilde{\chi}_0}^2 \log\left(\frac{E_R^{\mathrm{max}}}{E_R^{\mathrm{min}}}\right)  \right).
  \label{eq:Prim_e}
\ee

\bibliography{main}

\end{document}